\documentclass[11pt,twoside,a4paper]{cernrep} 
\usepackage{rep_common}
\usepackage{cancel}
\usepackage{placeins}
\usepackage{booktabs}
\usepackage{slashbox}
\usepackage{subfig}
\usepackage{amsmath,amssymb,amsthm}
\usepackage{rotating}
\usepackage{multicol}
\usepackage{tabulary}
\usepackage[colorlinks, urlcolor=blue, linkcolor=blue, citecolor=blue, plainpages=false]{hyperref}
\usepackage{float}
\floatstyle{plaintop}
\restylefloat{table}
\usepackage{xargs}
\usepackage{xstring}
\usepackage{lineno}

\usepackage{graphicx,multirow,colortbl}

\usepackage{fancyhdr}
\usepackage{siunitx}
\usepackage{textgreek}

\begin{document}
\newcounter{ncontacts}
\newcommand{\fcontact}[1]{\StrCount{#1}{,}[\tmp]\setcounter{ncontacts}{\tmp}
  Contact\ifthenelse{\value{ncontacts} > 0}{s}{}: #1}
\newcommand{\feditor}[1]{\StrCount{#1}{,}[\tmp]\setcounter{ncontacts}{\tmp}
  Contact Editor\ifthenelse{\value{ncontacts} > 0}{s}{}: #1}
\newcommandx{\asection}[2][1=NONE]{
  \ifthenelse{\equal{#1}{NONE}}
  {\section{#2}}{\section[#2]{#2\footnote{\feditor{#1}}}}}
\newcommandx{\asubsection}[2][1=NONE]{
  \ifthenelse{\equal{#1}{NONE}}
             {\subsection{#2}}{
               \subsection[#2]{#2\footnote{\fcontact{#1}}}}}
\newcommandx{\asubsubsection}[2][1=NONE]{
  \ifthenelse{\equal{#1}{NONE}}
             {\subsubsection{#2}}{
               \subsubsection[#2]{#2\footnote{\fcontact{#1}}}}}
\newcommandx{\asubsubsubsection}[2][1=NONE]{
  \ifthenelse{\equal{#1}{NONE}}
             {\subsubsubsection{#2}}{
               \subsubsubsection[#2]{#2\footnote{\fcontact{#1}}}}}

%Snowmass banner
\newcommand\snowmass{\begin{center}\rule[-0.2in]{\hsize}{0.01in}\\\rule{\hsize}{0.01in}\\
\vskip 0.1in Submitted to the  Proceedings of the US Community Study\\ 
on the Future of Particle Physics (Snowmass 2021)\\ 
\rule{\hsize}{0.01in}\\\rule[+0.2in]{\hsize}{0.01in} \end{center}}

% Title page.
\title{{\normalfont\bfseries\boldmath\huge
\begin{center}
A Muon Collider Facility for Physics Discovery
%
%\\[-10pt] {\normalsize{(Preliminary %Draft)}}
%
\end{center}%\vspace*{-30pt}
\vspace*{-40pt}
}
{\textnormal{\normalsize \snowmass}}
{\textnormal{\normalsize
This is one of the six reports submitted to Snowmass by the International Muon Collider Collaboration. \\
The Indico subscription page:\\
\centerline{
\href{https://indico.cern.ch/event/1130036/}{https://indico.cern.ch/event/1130036/}}
contains the link to the reports and gives the possibility to subscribe to the papers. \\
The policy for signatures is that, for each individual report, you can subscribe as "Author" or as "Signatory", defined as follows:
\begin{itemize}
    \item 
``Author'' indicates that you did contribute to the results documented in the report in any form, including e.g. by participating to the discussions of the community meeting, sending comments on the drafts, etc, or that you plan to contribute to the future work. The ``Authors'' will appear as such on arXiv.
\item
"Signatory means that you express support to the Collaboration effort and endorse the Collaboration plans. The "Signatories" list will be reported in the text only. 
\end{itemize}
The white papers will appear on arXiv on 15 March 2022 (Snowmass deadline).\\ Additional contributors can subscribe until March 30th and they will be added to the revised version.
}}
%{\textnormal{\normalsize \snowmass}}
}

\newcounter{instituteref}
\newcommand{\iinstitute}[2]{\refstepcounter{instituteref}\label{#1}$^{\ref{#1}}$\href{http://inspirehep.net/record/#1}{#2}}
\newcommand{\iauthor}[3]{\href{http://inspirehep.net/record/#1}{#2}$^{#3}$}
%%%%%%%%%%%%%%%%%%%%%%%%%%%%%%%%%%%%%%%%%%%%%%%%%%%%%%%%%%%
\author{Editors: \\ 
\iauthor{1058749}{D.~Stratakis}{\ref{902796}}, 
\iauthor{997065}{N.~Mokhov}{\ref{902796}}, 
\iauthor{994435}{M.~Palmer}{\ref{902689}}, 
\iauthor{994095}{{N.~Pastrone}}{\ref{902889}}, 
\iauthor{992212}{T.~Raubenheimer}{\ref{903206}}, 
\iauthor{1054170}{C.~Rogers}{\ref{903174}}, 
\iauthor{989645}{D.~Schulte}{\ref{902725}}, 
\iauthor{988978}{V.~Shiltsev}{\ref{902796}}, 
\iauthor{1268914}{J.~Tang}{\ref{1283410},\ref{903123}}, 
\iauthor{982905}{A.~Yamamoto}{\ref{902725},\ref{902916},} 
\\ \vspace*{4mm}Authors: \\ 
\iauthor{1757334}{C.~Aim{\`e}}{\ref{943385},\ref{902885}}, 
\iauthor{1067349}{{M.A.~Mahmoud}.}{\ref{912409}}, 
\iauthor{1073143}{N.~Bartosik}{\ref{902889}}, 
\iauthor{1037853}{{E.~Barzi}}{\ref{902796},\ref{903092}}, 
\iauthor{1049763}{A.~Bersani}{\ref{902881}}, 
\iauthor{1029828}{A.~Bertolin}{\ref{902884}}, 
\iauthor{1015872}{M.~Bonesini}{\ref{902882},\ref{907960}}, 
\iauthor{1670912}{B.~Caiffi}{\ref{902881}}, 
\iauthor{1057458}{M.~Casarsa}{\ref{902888}}, 
\iauthor{1014281}{M.G.~Catanesi}{\ref{902877}}, 
\iauthor{1021757}{A.~Cerri}{\ref{1241166}}, 
\iauthor{1937290}{{C.~Curatolo}}{\ref{902882}}, 
\iauthor{}{{M.~Dam}}{\ref{902882}}, 
\iauthor{1076225}{H.~Damerau}{\ref{902725}}, 
\iauthor{1651018}{{E.~De~Matteis}}{\ref{907142}}, 
\iauthor{1012025}{H.~Denizli}{\ref{908452}}, 
\iauthor{1031269}{B.~Di~Micco}{\ref{906528},\ref{907692}}, 
\iauthor{1011508}{T.~Dorigo}{\ref{902884}}, 
\iauthor{1069878}{S.~Farinon}{\ref{902881}}, 
\iauthor{1009979}{F.~Filthaut}{\ref{903075}}, 
\iauthor{1719039}{D.~Fiorina}{\ref{902885}},
\iauthor{1075917}{D.~Giove}{\ref{907142}},
\iauthor{}{M.~Greco}{\ref{906528}}, 
\iauthor{1007486}{C.~Grojean}{\ref{902770},\ref{902858}}, 
\iauthor{1006825}{T.~Han}{\ref{903130}}, 
\iauthor{1028687}{{S.~Jindariani}}{\ref{902796}}, 
\iauthor{1019544}{P.~Koppenburg}{\ref{903832}}, 
\iauthor{1252769}{K.~Krizka}{\ref{902953}}, 
\iauthor{1074984}{Q.~Li}{\ref{903603}}, 
\iauthor{1256188}{Z.~Liu}{\ref{903010}}, 
\iauthor{999862}{K.R.~Long}{\ref{902868},\ref{903174}}, 
\iauthor{999654}{D.~Lucchesi}{\ref{903113},\ref{902884}}, 
\iauthor{1670119}{{S.~Mariotto}}{\ref{903009},\ref{907142}}, 
\iauthor{1074063}{F.~Meloni}{\ref{902770}}, 
\iauthor{1461119}{C.~Merlassino}{\ref{903112}}, 
\iauthor{997679}{E.~M\'etral}{\ref{902725}}, 
\iauthor{2049482}{A.~Montella}{\ref{902888}}, 
\iauthor{1064125}{R.~Musenich}{\ref{902881}}, 
\iauthor{1069385}{M.~Nardecchia}{\ref{903168},\ref{902887}}, 
\iauthor{}{F.~Nardi}{\ref{903113},\ref{902884}}, 
\iauthor{995826}{{D.~Neuffer}}{\ref{902796}}, 
\iauthor{1048820}{S.~Pagan~Griso}{\ref{902953}}, 
\iauthor{1050691}{K.~Potamianos}{\ref{903112}}, 
\iauthor{1651162}{{M.~Prioli}}{\ref{907142}}, 
\iauthor{992463}{E.~Radicioni}{\ref{902877}}, 
\iauthor{992031}{L.~Reina}{\ref{902803}}, 
\iauthor{1020819}{C.~Riccardi}{\ref{943385},\ref{902885}}, 
\iauthor{1028713}{L.~Ristori}{\ref{902796}}, 
\iauthor{991185}{L.~Rossi}{\ref{903009},\ref{907142}}, 
\iauthor{990505}{P.~Salvini}{\ref{902885}}, 
\iauthor{990367}{J.~Santiago}{\ref{909079},\ref{903836}}, 
\iauthor{1019799}{A.~Senol}{\ref{908452}}, 
\iauthor{1342183}{L.~Sestini}{\ref{902884}}, 
\iauthor{1066476}{M.~Sorbi}{\ref{903009},\ref{907142}}, 
\iauthor{1319078}{G.~Stark}{\ref{1218068}}, 
\iauthor{1057643}{M.~Statera}{\ref{907142}}, 
\iauthor{1071725}{X.~Sun}{\ref{1210798}}, 
\iauthor{1265350}{I.~Vai}{\ref{902885}}, 
\iauthor{2025179}{R.~U.~Valente}{\ref{907142}} 
\\ \vspace*{4mm} Signatories: \\ 
\iauthor{1018902}{K.~Agashe}{\ref{902990}}, 
\iauthor{1018633}{B.C.~Allanach}{\ref{907623}}, 
\iauthor{1049113}{A.~Apresyan}{\ref{902796}}, 
\iauthor{1491320}{P.~Asadi}{\ref{1237813}}, 
\iauthor{}{D.~~Athanasakos}{\ref{910429}}, 
\iauthor{1041900}{A.~Azatov}{\ref{4416},\ref{902888}}, 
\iauthor{2031609}{F.~Batsch}{\ref{902725}}, 
\iauthor{1037456}{M.E.~Biagini}{\ref{902807}}, 
\iauthor{1020223}{K.M.~Black}{\ref{903349}}, 
\iauthor{1794682}{{.~~.~Bottaro}}{\ref{903128},\ref{902886}}, 
\iauthor{1015478}{{A.~Braghieri}}{\ref{902885}}, 
\iauthor{1015214}{A.~Bross}{\ref{902796}}, 
\iauthor{1894439}{L.~Buonincontri}{\ref{902884},\ref{903113}}, 
\iauthor{1077579}{D.~Buttazzo}{\ref{902886}}, 
\iauthor{1014742}{G.~Calderini}{\ref{926589},\ref{903119}}, 
\iauthor{1707397}{{S.~Calzaferri}}{\ref{902885}}, 
\iauthor{1024602}{A.~Canepa}{\ref{902796}}, 
\iauthor{1275234}{R.~Capdevilla}{\ref{908474},\ref{903282}}, 
\iauthor{}{L.~Castelli}{\ref{903113}}, 
\iauthor{1793525}{C.~Cesarotti}{\ref{902835}}, 
\iauthor{1037833}{G.~Chachamis}{\ref{1294609}}, 
\iauthor{1014143}{Z.~Chacko}{\ref{902990}}, 
\iauthor{2023221}{A.~Chanc\'e}{\ref{912490}}, 
\iauthor{2037614}{S.~Chen}{\ref{1471035}}, 
\iauthor{1069708}{Y.-T.~Chien}{\ref{1275736}}, 
\iauthor{1013275}{A.~Colaleo}{\ref{902660},\ref{902877}}, 
\iauthor{1862239}{M.~Costa}{\ref{903128},\ref{902886}}, 
\iauthor{1046385}{N.~Craig}{\ref{903307}}, 
\iauthor{1035631}{Y.~Cui}{\ref{903304}}, 
\iauthor{1024481}{D.~Curtin}{\ref{903282}}, 
\iauthor{1067364}{R.~T.~D'Agnolo}{\ref{1087875}}, 
\iauthor{2049478}{{G.~Da~Molin}}{\ref{909099}}, 
\iauthor{1012395}{S.~Dasu}{\ref{903349}}, 
\iauthor{1021004}{J.~de~Blas}{\ref{903836}}, 
\iauthor{1012237}{A.~Deandrea}{\ref{1743848}}, 
\iauthor{1012143}{J.~Delahaye}{\ref{902725}}, 
\iauthor{1019723}{A.~Delgado}{\ref{903085}}, 
\iauthor{1011983}{R.~Dermisek}{\ref{902874}}, 
\iauthor{1395010}{K.~F.~Di~Petrillo}{\ref{902796}}, 
\iauthor{1246709}{J.~Dickinson}{\ref{902796}}, 
\iauthor{1011443}{M.~Dracos}{\ref{911366}}, 
\iauthor{1404358}{F.~Errico}{\ref{902660},\ref{902877}}, 
\iauthor{1064320}{P.~Everaerts}{\ref{903349}}, 
\iauthor{1010523}{L.~Everett}{\ref{903349}}, 
\iauthor{1010065}{G.~Ferretti}{\ref{902825}}, 
\iauthor{1894571}{{M.~Forslund}}{\ref{910429}}, 
\iauthor{1052115}{{R.~Franceschini}}{\ref{906528},\ref{907692}}, 
\iauthor{1009120}{E.~Gabrielli}{\ref{903287},\ref{902888}}, 
\iauthor{1946817}{F.~Garosi}{\ref{904416}}, 
\iauthor{1894454}{L.~Giambastiani}{\ref{1513358},\ref{902884}}, 
\iauthor{1971617}{{C.~Giraldin}}{\ref{903113}}, 
\iauthor{1706734}{A.~Glioti}{\ref{1471035}}, 
\iauthor{1059457}{L.~Gray}{\ref{902796}}, 
\iauthor{1198373}{A.~Greljo}{\ref{902668}}, 
\iauthor{1274618}{J.~Gu}{\ref{903628}}, 
\iauthor{1007092}{H.E.~Haber}{\ref{1218068}}, 
\iauthor{1259916}{C.~Han}{\ref{903702}}, 
\iauthor{2044726}{J.~Hauptman}{\ref{902893}}, 
\iauthor{1383268}{B.~Henning}{\ref{1471035}}, 
\iauthor{1912097}{{K.~Hermanek}}{\ref{902874}}, 
\iauthor{1006149}{M.~Herndon}{\ref{903349}}, 
\iauthor{1067690}{T.R.~Holmes}{\ref{1623978}}, 
\iauthor{1515880}{S.~Homiller}{\ref{902835}}, 
\iauthor{1475406}{S.~Jana}{\ref{902841}}, 
\iauthor{2049476}{H.~Jia}{\ref{903349}}, 
\iauthor{1051663}{Y.~Kahn}{\ref{902867}}, 
\iauthor{1003695}{D.~M.~Kaplan}{\ref{1273767}}, 
\iauthor{1002991}{W.~Kilian}{\ref{903203}}, 
\iauthor{1020007}{K.~Kong}{\ref{902912}}, 
\iauthor{1345391}{G.K.~Krintiras}{\ref{1273495}}, 
\iauthor{1077491}{G.~Krnjaic}{\ref{902796}}, 
\iauthor{999784}{R.~LOSITO}{\ref{902725}}, 
\iauthor{1071846}{L.~Lee}{\ref{1623978}}, 
\iauthor{1064657}{W.~Li}{\ref{903156}}, 
\iauthor{999724}{I.~Low}{\ref{902645},\ref{903083}}, 
\iauthor{1700371}{{Q.~Lu}}{\ref{902835}}, 
\iauthor{1514492}{Y.~Ma}{\ref{903130}}, 
\iauthor{999053}{F.~Maltoni}{\ref{910783},\ref{902674}}, 
\iauthor{1668860}{L.~Mantani}{\ref{907623}}, 
\iauthor{1078065}{D.~Marzocca}{\ref{902888}}, 
\iauthor{1971307}{P.~Mastrapasqua}{\ref{910783}}, 
\iauthor{998430}{K.~Matchev}{\ref{902804}}, 
\iauthor{1054925}{A.~Mazzacane}{\ref{902796}}, 
\iauthor{1751811}{N.~McGinnis}{\ref{903290}}, 
\iauthor{1025277}{P.~Meade}{\ref{910429}}, 
\iauthor{997877}{{B.~Mele}}{\ref{902887}}, 
\iauthor{1022138}{P.~Merkel}{\ref{902796}}, 
\iauthor{1032624}{F.~Mescia}{\ref{905190},\ref{911212}}, 
\iauthor{1066275}{R.~K.~Mishra}{\ref{902835}}, 
\iauthor{1070072}{A.~Mohammadi}{\ref{903349}}, 
\iauthor{}{R.~Mohapatra}{\ref{99999}}, 
\iauthor{996989}{P.~Montagna}{\ref{943385},\ref{902885}}, 
\iauthor{995827}{N.~Neufeld}{\ref{902725}}, 
\iauthor{995794}{H.~Newman}{\ref{902711}}, 
\iauthor{995460}{{Y.~Nomura}}{\ref{903299}}, 
\iauthor{1070110}{I.~Ojalvo}{\ref{16750}}, 
\iauthor{1077958}{P.~Panci}{\ref{903129},\ref{902886}}, 
\iauthor{1274353}{D.~Pagani}{\ref{902878}}, 
\iauthor{1023838}{P.~Paradisi}{\ref{1513358},\ref{902884}}, 
\iauthor{1772198}{A.~Pellecchia}{\ref{902660}}, 
\iauthor{1067962}{{A.~Perloff}}{\ref{902748}}, 
\iauthor{993440}{F.~Piccinini}{\ref{902885}}, 
\iauthor{1021028}{M.~Pierini}{\ref{902725}}, 
\iauthor{1024769}{M.~Procura}{\ref{903326}}, 
\iauthor{1217056}{R.~Radogna}{\ref{902660},\ref{902877}}, 
\iauthor{992212}{T.~Raubenheimer}{\ref{903206}}, 
\iauthor{1214912}{D.~Redigolo}{\ref{1214912}}, 
\iauthor{1021811}{J.~Reuter}{\ref{902770}}, 
\iauthor{1056642}{F.~Riva}{\ref{902813}}, 
\iauthor{1040385}{T.~Robens}{\ref{902678}}, 
\iauthor{1066143}{F.~S.~Queiroz}{\ref{411233},\ref{411233}}, 
\iauthor{1072232}{F.~Sala}{\ref{908583}}, 
\iauthor{1885424}{J.~Salko}{\ref{902668}}, 
\iauthor{1077871}{E.~Salvioni}{\ref{1513358},\ref{902884}}, 
\iauthor{}{I.~Sarra}{\ref{99999}}, 
\iauthor{989950}{J.~Schieck}{\ref{903324},\ref{904536}}, 
\iauthor{1039590}{M.~Selvaggi}{\ref{902725}}, 
\iauthor{1071696}{V.~Sharma}{\ref{903349}}, 
\iauthor{1622677}{F.~M.~Simone}{\ref{902660},\ref{902877}}, 
\iauthor{}{A.~Stamerra}{\ref{902660},\ref{902877}}, 
\iauthor{1023964}{D.~Stolarski}{\ref{906105}}, 
\iauthor{1071880}{J.~Stupak}{\ref{1273509}}, 
\iauthor{987285}{S.~Su}{\ref{902647}}, 
\iauthor{987128}{{R.~Sundrum}}{\ref{912511}}, 
\iauthor{1078570}{M.~Swiatlowski}{\ref{903290}}, 
\iauthor{1454316}{A.~Sytov}{\ref{905268}}, 
\iauthor{986860}{T.~Tait}{\ref{903302}}, 
\iauthor{1878399}{E.~A.~Thompson}{\ref{902770}}, 
\iauthor{1064514}{R.~Torre}{\ref{902881}}, 
\iauthor{985810}{L.~~Tortora}{\ref{907692},\ref{902725}}, 
\iauthor{1778841}{S.~Trifinopoulos}{\ref{902888}}, 
\iauthor{1613622}{M.~Valente}{\ref{903290}}, 
\iauthor{1643523}{N.~Valle}{\ref{943385},\ref{902885}}, 
\iauthor{1071756}{R.~Venditti}{\ref{902660},\ref{902877}}, 
\iauthor{1063935}{P.~Verwilligen}{\ref{902877}}, 
\iauthor{1863232}{{L.~Vittorio}}{\ref{903128},\ref{902886}}, 
\iauthor{984555}{P.~~Vitulo}{\ref{943385},\ref{902885}}, 
\iauthor{1077733}{E.~~Vryonidou}{\ref{902984}}, 
\iauthor{1054127}{C.~Vuosalo}{\ref{903349}}, 
\iauthor{1073818}{{H.~Weber}}{\ref{902858}}, 
\iauthor{1511975}{{Y.~Wu}}{\ref{903094}}, 
\iauthor{1037622}{{A.~Wulzer}}{\ref{909099}}, 
\iauthor{1618109}{K.~Xie}{\ref{903130}}, 
\iauthor{1019845}{K.~Yonehara}{\ref{902796}}, 
\iauthor{1024759}{{H.-B.~Yu}}{\ref{903304}}, 
\iauthor{1971310}{A.~Zaza}{\ref{902660},\ref{902877}}, 
\iauthor{1066114}{Y.~J.~Zheng}{\ref{902912}}, 
\iauthor{1863481}{D.~Zuliani}{\ref{903113},\ref{902884}}, 
\iauthor{1037623}{J.~Zurita}{\ref{907907}} 
\vspace*{1cm}} \institute{\small 
\iinstitute{902796}{{Fermi National Accelerator Laboratory, Batavia, IL 60510-0500, United States}}; 
\iinstitute{902689}{{Department of Physics, Brookhaven National Laboratory, Upton, NY 11973-5000, United States}}; 
\iinstitute{902889}{{Istituto Nazionale di Fisica Nucleare, Sezione di Torino via P. Giuria, 1, 10125 Torino,  Italy}}; 
\iinstitute{903206}{SLAC National Accelerator Laboratory, United States}; 
\iinstitute{903174}{{Rutherford Appleton Laboratory, Harwell Oxford, Didcot OX110QX, United Kingdom}}; 
\iinstitute{902725}{CERN, Switzerland}; 
\iinstitute{1283410}{Sun Yat-sen University, guangzhou 510275, China}; 
\iinstitute{902916}{{High Energy Accelerator Research Organization KEK, Tsukuba, Ibaraki 305-0801, Japan}}; 
\iinstitute{903123}{Institute of High-Energy Physics, Beijing, China}; 
\iinstitute{943385}{Universit{\`a} di Pavia, Italy}; 
\iinstitute{912409}{{Center for High Energy Physics (CHEP-F), Fayoum University{\"a}, 63514 El-Fayoum, Egypt}}; 
\iinstitute{902881}{Istituto Nazionale di Fisica Nucleare - Sezione di Genova - via Dodecaneso 33, 16146 Genova - Italy}; 
\iinstitute{902884}{INFN Sezione di Padova, Padova, Italy}; 
\iinstitute{902882}{Istituto Nazionale di Fisica Nucleare, Italy}; 
\iinstitute{902888}{INFN Sezione di Trieste, Trieste, Italy}; 
\iinstitute{902877}{INFN , Bari, Italy}; 
\iinstitute{1241166}{MPS School, University of Sussex, Sussex House, BN19QH Brighton, United Kingdom}; 
\iinstitute{907142}{{Laboratori Acceleratori e Superconduttività Applicata (LASA), Istituto Nazionale di Fisica Nucleare (INFN), Via Fratelli Cervi 201, 20054 Segrate Milano, Italy}}; 
\iinstitute{908452}{Department of Physics, Bolu Abant Izzet Baysal University, 14280, Bolu, Turkey}; 
\iinstitute{906528}{Università degli Studi di Roma Tre, Italy}; 
\iinstitute{903075}{Radboud University and Nikhef, Nijmegen, The Netherlands}; 
\iinstitute{902885}{INFN, Sezione di Pavia, Pavia, Italy}; 
\iinstitute{902770}{{Deutsches Elektronen-Synchrotron DESY, Notkestr. 85, 22607 Hamburg, Germany}}; 
\iinstitute{903130}{University of Pittsburgh, United States}; 
\iinstitute{903832}{Nikhef National Institute for Subatomic Physics, Amsterdam, The Netherlands}; 
\iinstitute{902953}{{Physics Division, Lawrence Berkeley National Laboratory, Berkeley, CA, USA, United States}}; 
\iinstitute{903603}{Peking University, Beijing, China}; 
\iinstitute{903010}{{School of Physics and Astronomy, University of Minnesota, Minneapolis, MN 55455, USA}, United States}; 
\iinstitute{902868}{Imperial College London, Exhibition Road, London, SW7 2AZ, UK, United Kingdom}; 
\iinstitute{903113}{University of Padova, Italy}; 
\iinstitute{903009}{{Dipartimento di Fisica Aldo Pontremoli, Università degli Studi di Milano, Via Celoria, 16, 20133 Milano, Italy}}; 
\iinstitute{903112}{Particle Physics Department, University of Oxford, Denys Wilkinson Bldg., Keble Road, Oxford OX1 3RH, United Kingdom}; 
\iinstitute{903168}{Sapienza University of Rome, Italy}; 
\iinstitute{902803}{Florida State University, United States}; 
\iinstitute{909079}{{CAFPE}, Spain}; 
\iinstitute{1218068}{SCIPP, UC Santa Cruz, United States}; 
\iinstitute{1210798}{State Key Laboratory of Nuclear Physics and Technology, Peking University, Beijing, China}; 
\iinstitute{903092}{{Ohio State University, Columbus, OH 43210}, United States}; 
\iinstitute{907960}{Dipartimento di Fisica Universit\'a Milano Bicocca, Italy}; 
\iinstitute{907692}{Istituto Nazionale di Fisica Nucleare sezione di Roma Tre, Italy}; 
\iinstitute{902858}{{Humboldt-Universit\"at zu Berlin, Institut f\"ur Physik, Newtonstr. 15, 12489 Berlin, Germany}}; 
\iinstitute{902887}{INFN Rome, Italy}; 
\iinstitute{903836}{{Departamento de F\'isica Te\'orica y del Cosmos, Universidad de Granada, Campus de Fuentenueva, E--18071 Granada, Spain. }}; 
\iinstitute{902990}{Maryland Center for Fundamental Physics, University of Maryland, College Park, MD 20742, USA, United States}; 
\iinstitute{907623}{DAMTP, Wilberforce Road, University of Cambridge, Cambridge CB3 0WA, United Kingdom}; 
\iinstitute{1237813}{Center for Theoretical Physics, Massachusetts Institute of Technology,   Cambridge, MA 02139, USA., United States}; 
\iinstitute{910429}{YITP, Stony Brook, United States}; 
\iinstitute{4416}{SISSA International School for Advanced Studies, Via Bonomea 265, 34136, Trieste, Italy}; 
\iinstitute{902807}{INFN, Frascati National Laboratory, Italy}; 
\iinstitute{903349}{University of Wisconsin-Madison, United States}; 
\iinstitute{903128}{{Scuola Normale Superiore, Piazza dei Cavalieri 7, I-56126, Pisa, Italy}}; 
\iinstitute{902886}{INFN Sezione di Pisa, Largo B.\ Pontecorvo 3, 56127 Pisa, Italy}; 
\iinstitute{926589}{CNRS/IN2P3, Paris, France}; 
\iinstitute{908474}{Perimeter Institute, Canada}; 
\iinstitute{902835}{Department of Physics, Harvard University, Cambridge, MA, 02138, United States}; 
\iinstitute{1294609}{Laborat{\' o}rio de Instrumenta\c{c}{\~ a}o e F{\' \i}sica Experimental de Part{\' \i}culas (LIP),  Av. Prof. Gama Pinto, 2, P-1649-003 Lisboa, Portugal}; 
\iinstitute{912490}{IRFU, CEA, Université Paris-Saclay, Gif-sur-Yvette; France}; 
\iinstitute{1471035}{{Theoretical Particle Physics Laboratory (LPTP), Institute of Physics, EPFL, Lausanne, Switzerland}}; 
\iinstitute{1275736}{Physics and Astronomy Department, Georgia State University, Atlanta, GA 30303, U.S.A., United States}; 
\iinstitute{902660}{{Department of Physics, Universit{\`a} degli Studi di Bari, Italy}}; 
\iinstitute{903307}{University of California, Santa Barbara, United States}; 
\iinstitute{903304}{University of California-Riverside, United States}; 
\iinstitute{903282}{Department of Physics, University of Toronto, Canada}; 
\iinstitute{1087875}{Universit\`e Paris Saclay, CNRS, CEA, Institut de Physique Th\`eorique, 91191, Gif-sur-Yvette, France}; 
\iinstitute{909099}{{Dipartimento di Fisica e Astronomia, Universit\'a di Padova}, Italy}; 
\iinstitute{1743848}{IP2I, Universit\'e Lyon 1, CNRS/IN2P3, UMR5822, F-69622, Villeurbanne, France}; 
\iinstitute{903085}{University of Notre Dame, United States}; 
\iinstitute{902874}{Physics Department, Indiana University, Bloomington, IN, 47405, USA, United States}; 
\iinstitute{911366}{IPHC, Universit\'{e} de Strasbourg, CNRS/IN2P3, Strasbourg, France}; 
\iinstitute{902825}{Chalmers University of Technology 40220 Gothenburg , Sweden}; 
\iinstitute{903287}{Physics Department, University of Trieste, Strada Costiera 11, 34151 Trieste, Italy}; 
\iinstitute{904416}{SISSA, Italy}; 
\iinstitute{1513358}{Dipartimento di Fisica e Astronomia, Universit\`a degli Studi di Padova, Italy}; 
\iinstitute{902668}{Albert Einstein Center for Fundamental Physics, Institute for Theoretical Physics, University of Bern, CH-3012 Bern, Switzerland}; 
\iinstitute{903628}{Department of Physics, Fudan University, Shanghai 200438, China}; 
\iinstitute{903702}{School of Physics, Sun Yat-Sen University, Guangzhou 510275, China}; 
\iinstitute{902893}{{Iowa State University, Ames, Iowa,  50011 USA}, United States}; 
\iinstitute{1623978}{{University of Tennessee, Knoxville, TN, USA}, United States}; 
\iinstitute{902841}{Max-Planck-Institut f{\"u}r Kernphysik, Germany}; 
\iinstitute{902867}{{Department of Physics, University of Illinois at Urbana-Champaign, Urbana, IL 61801, USA}, United States}; 
\iinstitute{1273767}{Illinois Institute of Technology, United States}; 
\iinstitute{903203}{Department of Physics, University of Siegen, 57068 Siegen, Germany}; 
\iinstitute{902912}{Department of Physics and Astronomy, University of Kansas, Lawrence, KS 66045, USA, United States}; 
\iinstitute{1273495}{{Department of Physics and Astronomy 1082 Malott, 1251 Wescoe Hall Dr. Lawrence, KS 66045}, United States}; 
\iinstitute{903156}{Rice University, Houston, TX 77005, USA, United States}; 
\iinstitute{902645}{High Energy Physics Division, Argonne National Laboratory, Lemont, IL 60439, USA, United States}; 
\iinstitute{910783}{Center for Cosmology, Particle Physics and Phenomenology, Universit\'e catholique de Louvain, B-1348 Louvain-la-Neuve, Belgium}; 
\iinstitute{902804}{Physics Department, University of Florida, Gainesville FL 32611 USA, United States}; 
\iinstitute{903290}{TRIUMF, 4004 Westbrook Mall, Vancouver, BC, Canada V6T 2A3}; 
\iinstitute{905190}{Universitat de Barcelona, Spain}; 
\iinstitute{99999}{University of Maryland, college Park, USA, United States}; 
\iinstitute{902711}{California Institute of Technology, United States}; 
\iinstitute{903299}{{Department of Physics, University of California, Berkeley, CA 94720, USA}, United States}; 
\iinstitute{16750}{Princeton University, United States}; 
\iinstitute{903129}{Pisa University, Italy}; 
\iinstitute{902878}{INFN, Sezione di Bologna, Via Irnerio 46, 40126 Bologna, Italy}; 
\iinstitute{902748}{{Department of Physics, University of Colorado, 390 UCB, Boulder, CO 80309, United States}}; 
\iinstitute{903326}{University of Vienna, Faculty of Physics, Boltzmanngasse 5, 1090 Vienna, Austria}; 
\iinstitute{1214912}{INFN Florence, Italy}; 
\iinstitute{902813}{D\'epartment de Physique Th\'eorique, Universit\'e de Gen\`eve, 24 quai Ernest-Ansermet, 1211 Gen\`eve 4, Switzerland}; 
\iinstitute{902678}{Rudjer Boskovic Institute, Zagreb, Croatia}; 
\iinstitute{411233}{International Institute of Physics, Universidade Federal do Rio Grande do Norte, Campus Universitario, Lagoa Nova, Natal-RN 59078-970, Brazil}; 
\iinstitute{908583}{Laboratoire de Physique Th\'eorique et Hautes \'Energies, Sorbonne Universit\'e, CNRS, Paris, France}; 
\iinstitute{903324}{Institut f\"ur Hochenergiephysik der \"Osterreichischen Akademie der Wissenschaften, Nikolsdorfer Gasse 18, A-1050 Wien, Austria}; 
\iinstitute{906105}{Ottawa-Carleton Institute for Physics, Carleton University, 1125 Colonel By Drive, Ottawa, ON, K1S 5B6, Canada}; 
\iinstitute{1273509}{University of Oklahoma, United States}; 
\iinstitute{902647}{University of Arizona, United States}; 
\iinstitute{912511}{{Maryland Center for Fundamental Physics, University of Maryland, College Park, MD 20742, USA}, United States}; 
\iinstitute{905268}{INFN Division of Ferrara, Italy}; 
\iinstitute{903302}{Department of Physics and Astronomy, University of California, Irvine, CA 92697 US, United States}; 
\iinstitute{902984}{University of Manchester, Manchester M13 9PL, UK, United Kingdom}; 
\iinstitute{903094}{{Department of Physics, Oklahoma State University, Stillwater, OK, 74078, USA}, United States}; 
\iinstitute{907907}{{Instituto de F{\'i}sica Corpuscular, CSIC-Universitat de Val{\'e}ncia, Valencia, Spain}}; 
\iinstitute{903119}{LPNHE, Sorbonne Universit\'e, France}; 
\iinstitute{903083}{Department of Physics and Astronomy, Northwestern University, Evanston, IL 60208, USA, United States}; 
\iinstitute{902674}{Dipartimento di Fisica e Astronomia, Università di Bologna,  via Irnerio 46,  I-40126 Bologna  , Italy}; 
\iinstitute{911212}{Institut de Ciencies del Cosmos (ICC), Spain}; 
\iinstitute{904536}{Atominstitut, Technische Universit\"at Wien, Stadionallee 2,  A-1020 Wien, Austria}
} 

\begin{titlepage}

% Header ---------------------------------------------------
\vspace*{-1.8cm}

\noindent
\begin{tabular*}{\linewidth}{lc@{\extracolsep{\fill}}r@{\extracolsep{0pt}}}
\vspace*{-1.2cm}\mbox{\!\!\!\includegraphics[width=.14\textwidth]{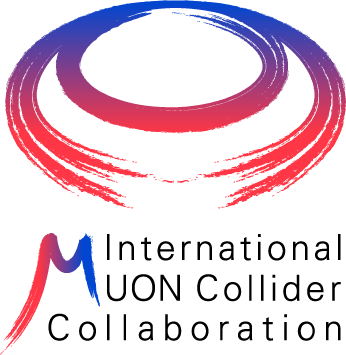}} & &  \\
 & & \today \\  % ID 
 & & \href{https://muoncollider.web.cern.ch}{https://muoncollider.web.cern.ch} \\ % Date - Can also hardwire e.g.: 23 March 2010
 & & \\
\hline
\end{tabular*}
\vspace*{0.3cm}
%
%\vspace*{2.0cm}
%
\maketitle

%\Title{The physics case of a 3~TeV muon collider stage}
%\vspace{\fill}
\newpage
% Abstract ------
\begin{abstract}
Muon colliders provide a unique route to deliver high energy collisions that enable discovery searches and precision measurements to extend our understanding of the fundamental laws of physics. The muon collider design aims to deliver physics reach at the highest energies with costs, power consumption and on a time scale that may prove favorable relative to other proposed facilities.  In this context, a new international collaboration has formed to further extend the design concepts and performance studies of such a machine.  This effort is focused on delivering the elements of a $\sim$10 TeV center of mass (CM) energy design to explore the physics energy frontier.  The path to such a machine may pass through lower energy options. Currently a 3 TeV CM stage is considered. Other energy stages could also be explored, e.g. an s-channel Higgs Factory operating at 125 GeV CM. We describe the status of the R\&D and design effort towards such a machine and lay out a plan to bring these concepts to maturity as a tool for the high energy physics community.
\end{abstract}

\vspace*{0.0cm}
\vspace{\fill}

\end{titlepage}

\section{Executive Summary}
High-energy lepton colliders can serve as facilities for precision and
discovery physics. The decrease of s-channel cross sections as $1/s$ requires
that luminosity increases with energy, ideally proportional to $s$, the square
of the centre-of-mass energy. The only mature technology to reach high-energy,
high-luminosity lepton collisions is linear electron-positron colliders; the
highest energy for which a conceptual design exists is the Compact Linear
Collider (CLIC) at 3 TeV.

The Muon Collider (MuC) promises to be able to extend
the lepton collider energy reach to much higher energies.
The strong suppression of synchrotron radiation compared to electrons allows
muon beam acceleration in rings making efficient use of the RF systems for
acceleration. The overall power consumption of a 10 TeV MuC is expected to be
lower than that of CLIC at 3 TeV. Additionally the beam can repeatedly produce
luminosity in two detectors in the collider ring. The ratio of luminosity to
beam power is expected to improve with collision energy, a unique feature of
the MuC. The compactness of the collider makes it plausible that a cost
effective design might be achieved; however this must be verified with more
detailed estimates. If the technical challenges can be overcome, the MuC offers
a potential route to long-term sustainability of collider physics.

To foster the Muon Collider concept an International Muon Collider
Collaboration (IMCC) has been initated after the recommendation of the update of the European Strategy for Particle Physics, initially hosted at CERN.  The
collaboration will address the muon collider challenges and develop the
concept and technologies in the coming years in order to be able to gauge if
the investment into a full conceptual design and demonstration programme is
scientifically justified. This will allow the Strategy Processes in the
different regions to make informed decisions.

Currently, the limit of the energy reach has not been identified. The study
focuses on a 10 TeV design with an integrated luminosity goal of 
$10 ~ \rm ab^{-1}$.
This goal is expected to provide a good balance between an excellent physics
case and affordable cost, power consumption and risk. Once a robust design
has been established at 10 TeV ~-~ including an estimate of the cost, power
consumption and technical risk ~-~ other, higher energies will be explored.

A potential initial energy stage at 3 TeV with an
integrated luminosity of $1 ~ ab^{-1}$ is also considered. 
This option might cost
around half as much as the 10 TeV option, and can be upgraded to 10 TeV or
beyond by adding an accelerator ring and building a new collider ring. Only
the 4.5 km-long 3 TeV collider ring would not be reused in this case.
This stage could potentially start colliding beams in the mid
2040s ~-~ depending on the strategic decisions.
This also requires that sufficient funding
is available already during the design phase and that all challenges can be
successfully addressed with no delays.

The muon collider thus promises a sustainable path toward the
very high energy. Its large energy and luminosity reach makes the direction
and the development of the technologies attractive. Potential intermedidate
stages provide important physics results on the way on timescales more adapted
to the human life span and provide the motivation for the scientists and
engineers that is the most important driver of the technological progress.

Muon Collider technology must overcome several
significant challenges to reach a level of maturity similar to linear
colliders. An increased
level of R\&D effort is justified at the current time, because the muon
collider promises an alternative path toward high-energy, high-luminosity
lepton collisions that extends beyond the expected reach of linear colliders. Supporting technologies such as high-power proton drivers, high-field solenoids and high-gradient RF cavities have, in the last decade, approached the level required to deliver the requisite luminosity.

Past work has demonstrated several key MuC technologies and concepts, and gives
confidence that the concept is viable. Component designs have been developed
that can cool the initially diffuse beam and accelerate it to multi-TeV energy
on a time scale compatible with the muon lifetime. However, a fully integrated
design has yet to be developed and further development and demonstration of
technology are required. In order to enable the next European Strategy for
Particle Physics Update (ESPPU), the next Particle Physics Project
Prioritisation Process (P5) and other
strategy processes to judge the scientific justification of a full
Conceptual Design Report (CDR) and demonstration programme, the design and
potential performance of the facility must be developed in the next few
years.

An R\&D programme has been developed in the frame of the European Roadmap for
Accelerator R\&D. The programme addresses the key challenges and is based
on consultations of the community at large, combined with the expertise of a
dedicated Muon Beams Panel. It benefited from significant input from the MAP design and studies
and US experts.

The proposed programme of work,
if fully resourced, % CTR
%~-~ in the so called aspirational scenario ~-~
will
allow the assessment of realistic luminosity targets, detector backgrounds,
power consumption and cost scale,
as well as whether one can consider implementing a MuC at CERN or elsewhere.
Mitigation strategies for the key technical risks and a demonstration
programme for the CDR phase will also be addressed.
The use
of existing infrastructure, such as existing proton facilities and the LHC
tunnel, will also be considered. Based on the conclusions of the next strategy processes in the different regions, a CDR phase could then develop the technologies and the baseline design
%needed to mitigate identified project risks and 
to demonstrate that the community
can execute a successful MuC project.

%The work programme will
%develop a muon collider concept at 10 TeV and explore a 3 TeV staging to
%mitigate technology and operational challenges.

No cost estimate for the CDR and demonstrator phase exists but
experience indicates that typically 5-10\% of the final project cost would
need to be invested before the project construction can start. A muon cooling demonstrator facility would be expected to
be the largest single component of the CDR programme, with the potential to
provide direct scientific output in its own right.

The resources available to the MuC programme over the next five years will
depend on decisions made in the different regions and in particular in the US.

%; the strategy process in thelatter case will conclude in 2023. Currently, CERN plans a budget of 2 MCHF per year and several person-years have already been committed at INFN, allowing the work to start. 

The muon collider
programme is synergistic with other R\&D efforts, and will directly
benefit from progress in the domains of high-field magnets, RF systems and recirculating linacs. The muon collider requires high-field dipoles, similar to proton facilities, and in particular robust high-field solenoids. These have a strong synergy with the development for other fields of science and society, e.g. the development of fusion power, high-field science, NMR spectroscopy and MRI machines of the next generation.  

Bright muon
beams are also important for potential future neutrino physics facilities such as NuSTORM or
ENUBET. A muon cooling
demonstrator facility could potentially share a large part of the
infrastructure with these facilities, from the proton source to the target.

\section{Design Overview}

\subsection{Status of Design}
MAP has develop many concepts and technologies for the muon collider. The rate of progress in this field gives confidence that the muon collider is a viable option. However no integrated design design of the muon collider has been made and further development of the technologies is required. The goal in the coming years is to develop the concept and technologies to a level that one can justify the investment into a technology and concept demonstration programme, including the full concept development and optimisation. The goal of this programme is to be able to commit to the project and to prepare the technical design phase.

\subsection{Performance Matrix}

\subsubsection{Attainable Energy}
At this moment, no fundamental limit for the energy reach of the muon collider
has been identified. The limitations are expected to come from the cost of the
collider complex and from the technological and power consumption challenges
to obtain high luminosity at high energy. In this respect the energy reach cannot
be separated from the luminosity reach.

The goals for the muon collider design study are thus driven by a balance
between an excellent physics case and affordable cost, power consumption and
technical risk. The 10 TeV goal is chosen for these very reasons and supported
by a very strong physics case.

Once a firm design has been established for an energy of 10 TeV, including an
estimate of the scale of cost, power consumption and technical risk, other higher
energies will also be explored.

A 3 TeV design will also be studied as it may be an attractive entry stage for
the muon collider with a cost well below the 10 TeV and an already strong physics case.
For this concept the challenges in the high-energy complex are reduced; e.g. the requirements
for the final focus system magnets are in the same ballpark as for the the HL-LHC.
Because of its smaller budget need and size such a stage could be most likely be implemented
earlier than the 10 TeV stage. It could be then upgraded to 10 TeV by reusing the full
complex, except for the 4.5 km-long 3 TeV collider ring, and by adding an
additional accelerator ring and a new collider ring.

\subsubsection{Attainable Luminosity and Luminosity Integral}
The integrated luminosity targets for the muon collider are based on requirements from
physics. They follow a scaling to keep the number of $s$-channel events constant
at all energies.

The tentative parameters and luminosity targets have been chosen based on previous
MAP studies. They would achieve the integrated luminosity target within five years in
a single detector. Paths have been identified to achieve the parameters. The focus of
the study is to address the technological challenges on these paths to achieve
a robust performance prediction.

Alternative approaches that could lead to higher luminosity will also be
explored, e.g. Parametric Ionization Cooling. However, in the near term the focus
is to ensure a solid design and advancing technologies.

%\subsubsection{Injector and driver systems}

%\subsubsection{High-energy complex}

\subsubsection{Facility Scale}
The dimensions of the muon collider are not known and depend on future technology and design choices. Still, some indication of the dimensions can be
estimated. 

Considering the largest RCS and collider ring at
10 TeV, one can assume 16 T field in the superconducting static dipoles of
both rings and a field range of $\pm2\rm T$ for the fast-ramping normal magnets in the RCS.
In both cases the effective dipole filling factor could be 80\%, similar to the
LHC. This accounts for flanges, instrumentation, focusing quadrupoles, etc.
In case of combined function magnets the filling factor can be higher but the
dipole field might have to be reduced in proportion. Further some space for
the RF and injection, extraction and interaction region is required.
Based on this, the initial estimate of the length is 10 km for the 10 TeV
collider and 32 km for the pulsed synchrotron that accelerates the beam from
1.5 to 5 TeV. The total tunnel is comparable to a 3 TeV CLIC,
which has an estimated tunnel length of about 50 km to house
the linacs and the final focus systems. The pulsed synchrotron could be made
more compact using fast ramping HTS magnets with a larger field reach or by
performing the acceleration in two stages in the same tunnel, which would require
less than 20 km but more superconducting magnets.

The final pulsed synchrotron for the 3 TeV stage would be approximately 12 km
long and the collider 4.5 km, using superconducting dipoles with a field of
only 11 T.

It is also possible to consider using the LHC tunnel for the final pulsed
synchrotron of the 3 TeV stage and for the acceleration to 10 TeV. However
more detailed studies are required.

\subsubsection{Power requirements}
The goal is to remain at a power consumption for the 10 TeV
collider well below the level of CLIC at 3 TeV or FCC-hh. The design has to advance more to assess the power consumption scale in a robust fashion. The scaling of the power consumption with energy is illustrated by some tentative considerations below.

A number of key components drive the power consumption if one extends to high energy:
\begin{itemize}
\item The power loss in the fast-ramping magnets of the pulsed synchrotron and their power converter. This is being addressed by a dedicated study.
%The goal is to remain at less than 10 MJ of losses per pulse, i.e. less than 50 MW.
\item The cryogenics system that cools the superconducting
magnets in the collider ring. This depends on the efficiency of shielding the magnets from the muon decay-induced heating.
%One can conservatively assume that the magnets operate at 2 K. Static losses would be of the order of 1 W/m. The dynamic losses due to the debris from the muon beam decay amount to a total of 5 MW, if the beam fully decays. Based on previous studies one can expect 1\% to reach the magnets and the rest to remain in the warm shielding. The cryogenics to remove the total heat load of 60 kW requires 42 MW of power from the grid.
\item The cryogenics power to cool the superconducting magnets and RF cavities in the pulsed synchrotron.
%This includes static losses of 1 W/m and assumes that 90\% of the muon beam debris can be removed by collimation.
\item The power to provide the RF for accelerating cavities in the pulsed synchrotron.
\end{itemize}
%After subtracting the power consumption of the 3 TeV collider ring, one could thus estimate that the 10 TeV collider requires and additional 150 MW. All of these power drivers will be assessed in the next few years.
The first contribution requires a particular effort as it depends on unprecedented large-scale fast ramping systems and the second and third contributions depends on the shielding choices. The contributions can be estimated reliably once the design choices ~-~ such as magnet material and operating temperature, RF system design, shielding thickness etc. ~-~ have been made.

\subsection{Design Summary}

The current muon collider baseline concept was developed by the Muon Accelerator Program (MAP)
collaboration \cite{bib_muon:mapweb}, which conducted a focused program of technology R\&D to evaluate its feasibility.
Since the end of the MAP study seminal measurements have been performed by the Muon Ionization
Cooling Experiment (MICE) collaboration, which demonstrated the principle of ionisation cooling that
is required to reach sufficient luminosity for a muon collider \cite{bib_muon:muoncolliderweb}. The MAP scheme is based on the
use of a proton beam to generate muons from pions decay and is the baseline for the collider concept
being developed by the new international collaboration. An alternative approach Low Emittance Muon Accelerator (LEMMA), which uses positrons to produce muon pairs at threshold, has been explored at INFN\cite{bib_muon:LEMMA}.

\begin{figure}
        \includegraphics[width=0.632\textwidth, trim={2.1cm 0 0 0}]
        {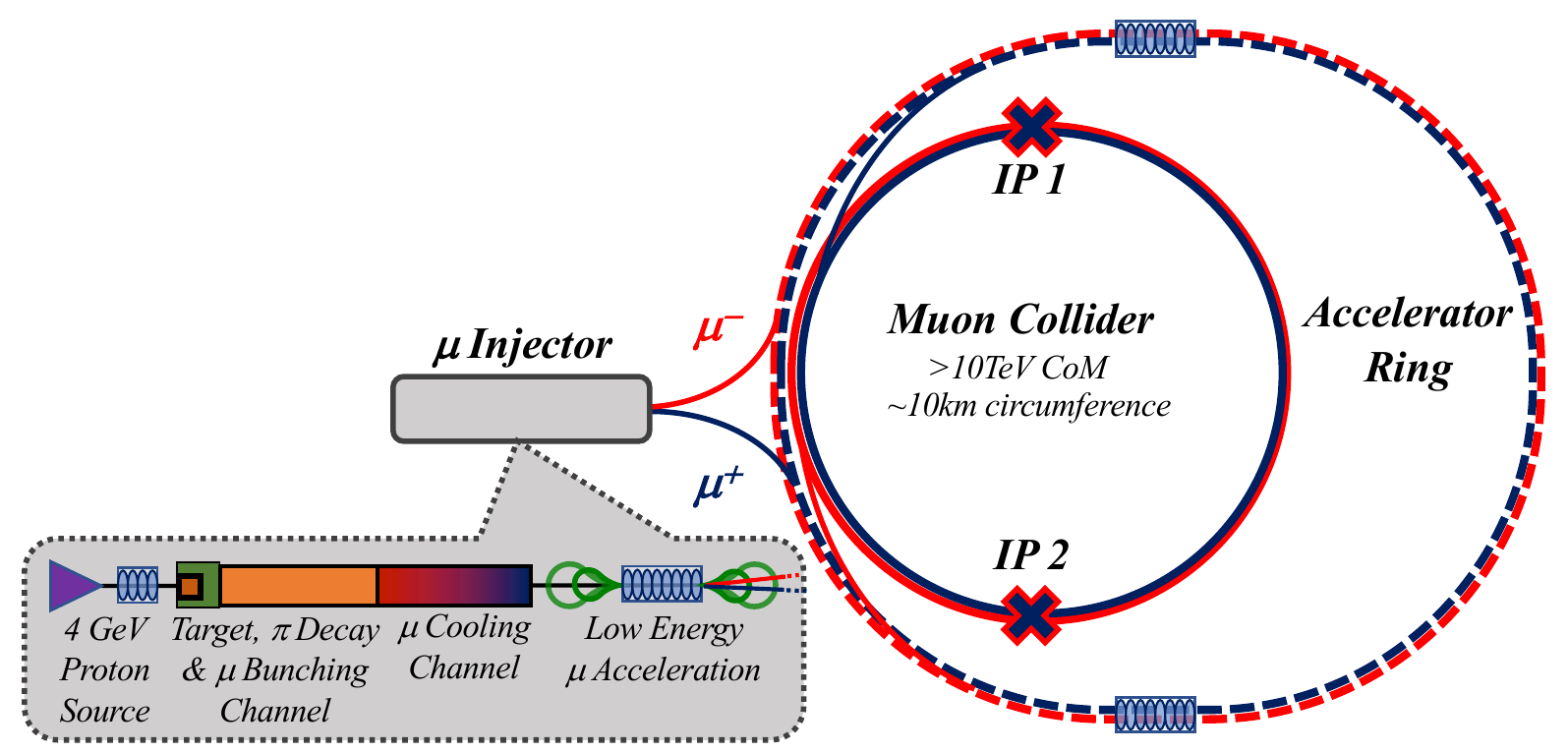}
        \centering\textbf{}

\caption{A conceptual scheme for the Muon Collider}
\label{fig_muon:sketch}
\end{figure}

MAP developed the concept shown in Fig.~\ref{fig_muon:sketch}. The proton complex produces a short, high intensity
proton pulse that hits a target and produces pions. The decay channel guides the pions and
collects the muons produced in their decay into a buncher and phase rotator system to form a muon
beam. Several cooling stages then reduce the longitudinal and transverse emittance of the beam using
a sequence of absorbers and RF cavities in a high magnetic field. A linac and two recirculating linacs
accelerate the beams to 60 GeV. One or more rings accelerate the beams to the final energy. As the
beam is accelerated, the lifetime in the lab frame increases due to relativistic time dilation so later stage
accelerators have proportionally more time for acceleration, so that fast-pulsed synchrotrons can be used.
Fixed-Field Alternating-gradient (FFA) accelerators are an interesting alternative. Finally the two single bunch
beams are injected at full energy into the collider ring to produce collisions at two interaction
points.
The MAP study demonstrated feasibility of key components, but several important elements were
not studied. The highest collision energy studied by MAP was 6 TeV. Technical limitations such as
beam-induced backgrounds have not been studied in detail at higher energies. Individual elements of the
muon source were studied, but integrated system design and optimisation was not performed. Cooling studies assumed limits in practical solenoid and RF fields that now appear to be too conservative; an
updated performance estimate would likely yield a better assessment of the ultimate luminosity of the facility.
MAP studies considered gallium, graphite and mercury target options, which should be progressed
and studied in more detail to assess fully the performance and technical limitations of the system.

\subsection{Design Challenges}

\subsubsection{Muon beam}

Although muons offer many potential physics benefits, their use brings substantial complications as well. Indeed, if intense muon beams were easy to produce, they would already be available. Firstly, muons are created as a tertiary beam. The proposed production scheme uses a proton beam to bombard a high-Z target. This produces pions, which are captured in a solenoidal decay channel, where they decay to muons. To produce an acceptably large sample of muons, a multi-MW proton beam is required.
%; a typical Neutrino Factory specification is for a MW scale proton driver. 
A target system capable of tolerating such an intense beam is a substantial challenge. The capture and decay process just described gives rise to a muon beam having a large energy spread and a large transverse phase space. The large transverse phase space has several implications: (1) it favors the use of solenoidal focusing in the lower energy portions of the facility, as opposed to the more conventional quadrupole focusing. A solenoid focuses in both planes simultaneously, avoiding the excessively large beam size in one plane when using an alternating polarity quadrupole channel. (2) it requires a rapid mechanism for reducing the emittance to more tractable values.
%; (3) it requires a high-acceptance fast acceleration system and decay ring. 
The second major challenge of muon beams is due to the short lifetime of the muon, only 2.2 $\mu$s at rest. Clearly, the short lifetime puts a premium on very rapid beam manipulations. A fast emittance cooling technique, “ionization cooling,” is needed to reduce the transverse emittance of the muon beam, along with a very rapid acceleration system. The ionization cooling technique requires high-gradient normal conducting RF cavities due to the need to immerse the cavities in a strong solenoidal magnetic field.
Finally, the decays of the muons lead to potentially severe backgrounds in the detector of a Muon Collider. There are also a number of challenges related to the magnet requirements.
In the target area, the initial capture magnet is a 20 T solenoid design. In the cooling channel, large aperture magnets having relatively low field, up to 2-3 T and 1.5-m diameter, are utilized. As the beam emittance is reduced through cooling, higher field solenoids with lower diameter bores are needed. In the final cooling stages of a Muon Collider, very high strength solenoids, up to $\sim$50 T, are required. In the acceleration system, solenoids with very low fringe fields are needed to permit operation of nearby superconducting RF cavities. In the acceleration system and collider ring, special split-midplane or shielded dipoles are needed to
accommodate the high heat load from muon decay electrons.

\subsubsection{Machine Detector Interface (MDI)}
A multi-TeV Muon Collider  needs to deliver high instantaneous luminosity ($\sim$10$^{34}$-10$^{35}$ cm$^{-2}$s$^{-1}$) to be able to exploit the diverse  physics potential opportunities. To reach this goal high intense muon beams, $2\times 10^{12}$ muons per bunch, are necessary. 
%na number of demanding requirements to the collider optics and to the Interaction Region (IR) design should be satisfied, arising from the short muon lifetime.
%While it is mandatory to protect from the muon decay products all superconducting magnets along the collider ring, 
At the Interaction Point (IP), where the two $\mu{^+}\mu{^-}$ bunches collide, the Beam-Induced Background (BIB) strongly affects the detectors' performance; therefore  an efficient mitigation strategy is required. The background arriving on the detector volume is a huge flux of secondary and tertiary low energy particles, mainly originated from muon decays occurring several decades of meters from the IP, with an additional component of incoherent e$^+$e$^-$ pair production and Bethe Heitler muons.
The current solution to mitigate such a background flux, initially proposed by MAP, consists of two tungsten cone-shaped shielding (nozzles) around the beampipe, with the origin in proximity of the interaction point, to be accurately designed and optimized for each specific beam energy.  
%To absorb the lowest energy neutrons the nozzles are coated with borated polyethylene. 
The design of the detector and the choice of the sub-detector technology need to cope with the nozzle structure and with the remaining fluxes of particles. The interaction region at 1.5 TeV center of mass energy has been designed and optimized by MAP. It was simulated with a framework developed at Fermilab \cite{bib_muon:MARS1} with the MARS15 code, and it is used as a starting point and benchmark. 
%A 750 GeV muon has a decay length  $\lambda = 4.7 \times 10^6 m$.  A bunch of $ 2 \times 10^12$ muons produces $4.28 \times 10^5$ decays per meter of lattice in a single pass. The main BIB components are neutrons, photons and electrons/positrons with low momentum, displaced origin and asynchronous time of arrival respect to the collision time.

Within the International Muon Collider Collaboration, a new framework, based on FLUKA,  has been developed to study the beam-induced background  at different energies\cite{bib_muon:fluka_bib}.
Studies performed so far demonstrate that, given reasonable assumptions of detector performance, it will be possible to perform the most challenging physics measurements\cite{bib_muon:detector}. 
%Optimisations, for example using improved pixel timing on the tracker detector and novel trigger algorithms, are in progress and may yield improved performance. This requires further studies at higher energies. 
Preliminary studies performed with a very preliminary configuration of the IR at 10 TeV center of mass energy with the nozzle designed for the 1.5 TeV, indicate that the beam induce background levels around the IP do not increase significantly.
Combined interaction region, detector shielding and detector design should be performed to confirm physics performance at 3~TeV and 10~TeV.

\subsubsection{Required key technologies}
% Sliced by subsystem rather than by technology
% Too detailed? Or did I forget anything?
The following technologies are required in order to deliver the muon collider.
\begin{itemize}
\item Development of a high power proton driver.
\begin{itemize}
\item Delivery of a high intensity H- ion source.
\item Stripping of the H- ions without excess foil heating.
\item A bunch compressor ring to deliver protons in a single, approximately 2 ns long bunch onto the target.
\end{itemize}
\item A target system capable of managing the large instantaneous beam power.
\begin{itemize}
\item A target that can withstand both high instantaneous current and high average beam power.
\item A suitable shielding system to protect the sensitive capture magnet that surrounds the target.
\item A superconducting capture magnet that can withstand radiation leakage from the target, that can contain the required shielding and that can deliver 15-20 T field to contain the pion beam. 
\item A beam dump that can capture the remnant proton beam without detrimental effect on the outgoing pion and muon beam.
\end{itemize}
\item A muon cooling system capable of delivering orders of magnitude of emittance reduction.
\begin{itemize}
\item A system of intermediate field solenoids, tightly integrated with RF cavities and absorbers.
\item A few extremely high-field, small-bore solenoids, with magnetic fields exceeding the current state of the art ($\sim$30~T).
\item Normal conducting RF systems capable of providing high field gradients in the presence of strong magnetic fields.
\end{itemize}
\item A beam acceleration system that can accelerate the muon beam on a time scale that is compatible with the time-dilated muon lifetime.
\begin{itemize}
\item Fast ramping power supplies to deliver ramp times of several T on a timescale of ms.
\item Superconducting RF systems robust to beam loading and capable of delivering significant RF gradient for rapid muon acceleration.
\end{itemize}
\item A collider ring that can deliver the collisions.
\begin{itemize}
\item Dipole magnets operating at high field, to recirculate the beam as many times as possible before the muons decay.
\item Dipole magnets must be robust to the radiation arising from muon decay.
\item The collider ring systems must be movable such that neutrino flux does not produce showering off-site from the collider ring.
\item High gradient final focus quadrupoles, comparable to HL-LHC at 3 TeV and more demanding at 10 TeV, that can deliver the small $\beta^*$ required to deliver requisite luminosity.
\item A detector system capable of distinguishing signal from the beam-induced background arising due to muon decays in the collider ring.
\end{itemize}
\end{itemize}

\subsubsection{Environmental impacts}
The muon collider will impact the environment in several ways. Key factors are:
\begin{itemize}
\item The land use for the construction of its infrastructure.
\item The power that the collider consumes.
\item The waste that it generates.
\item The production and transport of components.
\item Other impact on the environment, e.g. potential radiation.
\end{itemize}

\subsubsection{Land use}
As discussed above, the dimensions of the muon collider are not known and depend on future technology and design choices. Still, some indication of the dimensions can be estimated. 

The size for the last pulsed synchrotron and the collider ring could 32 and 10 km, respectively. The collider is therefore expected to be comparable to CLIC at 3 TeV in terms of land use.

If it is possible to reuse the LHC tunnel for the pulsed synchrotron, the total amount of tunnel that needs to be
constructed would be significantly reduced. This option will be explored.

\subsubsection{Power consumption}
The goal is to have a power consumption of the 10 TeV stage significantly
below the one of CLIC at 3 TeV or the FCC-hh. No detailed study has been
performed up to now. MAP estimated a power consumption of roughly 100 MW, independent of the colliding beam energy, for the muon production \cite{bib_muon:power}. We will
develop an improved estimate and focus on the improvement of power consumption
drivers.

\subsubsection{The generated waste}
The muon collider uses leptons in the high-energy part of the collider
and is thus a relatively clean machine. The main concern would be the
proton target. To minimise the impact of the target, we plan not to use
a mercury target but rather a solid graphite target. Other options
such a liquid gallium or fluidised tungsten will also be explored and are also
more environmentally sensitive than mercury.
The total power is in the range of existing and planned targets
at other facilities. Therefore no specific problem is expected.

\subsubsection{Production and transport of components}
This issue has not been studied. However the superconducting systems would be
comparatively compact, which indicates that the load to the environment is
comparatively limited.

\subsubsection{Other impact on the environment}
The proposed mechanical system can mitigate the neutrino flux density resulting
from muon decay from the arcs sufficiently even at 14 TeV. The insertions, in
particular for the experiments, complicate the mitigation and have to be
investigated. We expect to be able to find an orientation for the collider
complex that mitigates the potential issues.

%%% Local Variables:
%%% mode: latex
%%% TeX-master: t
%%% End:

\section {Technology Requirements}

\subsection {Technology Readiness Assessment}
A set of key challenges has been identified for the muon collider.
For each of these challenges a path toward addressing it has been identified.
Important R\&D remains to demonstrate that all obstacles on these paths can
be overcome.

It is important to note that in several areas the design targets are
indicative and failure to fully achieve them can be mitigated elswhere in the
design. To develop an integrated concept of the collider that allows to
fully assess the consequences of such trade-offs is an important part of the
studies.

The main risk for the muon collider design arises from insufficient resources
to address the challenges.

\subsubsection {Proton complex}
The challenge of the proton complex arises from the need to
combine the protons into short high-charge bunches. The proton source is site-dependent and designs for proton facilities compatible with a muon collider exist. Studies will be performed as resources allow. 

\subsubsection {High-power target and solenoid}
A high-power target is essential to produce a sufficient number of protons to
achieve the luminosity goal. Key challenges are the survival of the target
itself under the shock waves of the incoming beam pulses and the temperature
gradients to remove the deposited heat.

Compared to the MAP study, the power in the target is reduced by a factor
three. This allows to consider the use of a solid target
in addition to a liquid metal or a fluidized tungsten target. Unwanted particles 
produced by the proton beam impacting the target leads to heating
of the surrounding superconducting solenoid and to radiation damage.

Simulations of a 4 MW mercury-based target demonstrated that a 1.2 m radius of
the solenoid provides enough space to shield the magnet and to reach peak powers
in the coil of less than 0.1 mW/g, which corresponds to O(1 MGy) per year.
FCC-hh assumes that the magnet insulation can withstand an accumulated dose of
30 MGy\cite{bib_muon:fcc-hh}.
This solenoid is very demanding and resembles in cost and stored energy the
central solenoid of ITER. 

\subsubsection {Muon ionization cooling channel design}
Muon ionization cooling is required to increase the muon beam brightness to 
yield a sufficient luminosity. Ionization cooling requires the beam to be 
repeatedly slowed in absorbers and reaccelerated in RF cavities, while 
maintaining tight focusing of the muon beam.

The large longitudinal emittance of the beam arising from the target is captured 
in a sequence of RF cavities as a train of bunches. The emittance is reduced by a
rectilinear cooling system comprising weak dipole fields and increasingly strong
solenoids. The initial part of the system is optimised for large acceptance,
while the final part of the system is optimised for low emittance. When the beam
emittance is sufficiently reduced, the bunch train is merged into a single bunch
and then cooled further. 

In order to reach the lowest emittances, very strong solenoid fields are 
required with lower beam energy. In this regime longitudinal emittance increases and
a phase rotation system is required to minimise energy spread and control 
chromatic aberrations. Low frequency RF cavities or induction linacs, in regions 
of relatively low magnetic field, are used for reacceleration.

\subsubsection {Operation of RF cavities in a magnetic field}
In order to maintain the tight focusing required to yield good cooling 
performance in the rectilinear cooling system, a compact lattice is required 
with large real-estate RF gradient.
This results in a lattice that has RF operating near to the break down limit 
while immersed in a strong solenoid field.

The operation of RF cavities in a solenoid field poses specific challenges.
The solenoid field guides electrons that are emitted at one location of the
cavity surface to another location on the opposing wall and leads to localised
heating that can result in breakdown and cavity damage. Operation of copper
cavities is 3 T magnetic field showed a maximum useable gradient of only
10 MV/m.

Three approaches to overcome this obstacle are known:
\begin{itemize}
\item Use of lower-Z materials such as beryllium to limit the energy loss
  density.
\item The use of high-pressure hydrogen gas inside the cavity. In this case the 
mean free path of the electrons is limited and does not allow them to gain
  enough energy to ionise the gas or to produce a breakdown.
\item The use of very short RF pulses to limit the duration of the heat load in
  the cavity.
\end{itemize}
The first two techniques have been experimentally verified in MUCOOL with a
field of about 3 T (limited by the solenoid). They yielded a gradient of 50
MV/m in a beryllium cavity under vacuum and 65 MV/m in a molybdenum cavity
with hydrogen \cite{bib_muon:MAP_VacRF},  demonstrating no degradation in achievable field in the presence of an applied magnetic field.

Systematic studies in a new test stand are required to further develop the
technologies.
It will also be important to test the cavities in the actual field
configuration of the cooling cell, which differs from a homogeneous
longitudinal field.

\subsubsection {Final cooling and HTS magnet technology}
In the final muon cooling system, solenoids with the highest practical field
are needed. A design based on 30 T solenoids---a value that has been achieved
experimentally---demonstrated that an emittance about a factor two above the
target can be achieved \cite{bib_muon:MAP_FinalCooling}; it should be noted that the study aimed at this
larger target. Several options to improve the emittance will be
studied. With the technology progress, solenoids with field above 40 T are
being planned and would benefit the muon collider. Operating the cooling at
lower beam energy also allows to reach a smaller emittance; preliminary studies
indicate that 30 T may be marginally sufficient to reach the emittance target.

\subsubsection{Acceleration stage design}
The baseline acceleration stage consists of some linacs followed by a sequence
of pulsed synchrotrons that performs the lion share of the acceleration.
The synchrotrons combine static superconducting and fast-ramping magnets,
which could be either normal- or superconducting.
Fast-ramping magnets have been designed and models tested in the MAP study.
However, an integrated study of the lattice, the accelerating RF and the
fast-ramping magnet system including its powering system is essential to
estimate the cost and power consumption of the system.
The energy in the field of the magnets is of O(100 MJ). It needs to be
recovered from each pulse to the next with high efficiency. The requirements
on the quality of the ramp will drive cost and power consumption.
Studies in the coming years will address this challenge.

\subsubsection{Collider ring design}
The collider ring requires a small beta-function at the collision point,
resulting in significant chromaticity that needs to be compensated. It also
needs to maintain a short bunch. A solution for 3 TeV has been developed
and successfully addressed the challenges. A design of 10 TeV is one of the key 
ongoing efforts.

High-energy electrons and positrons arising from muon decay and striking the
collider ring magnets can cause radiation damage and unwanted heat load.
This can be mitigated with sufficient tungsten shielding; a successful design
has been developed at 3 TeV. First studies at 10 TeV indicate that the
effect is comparable to 3 TeV, since the power per unit length of the
particle loss remains similar.

The shielding requires a substantial aperture in the superconducting magnets.
The limit for the dipole field is thus given by the maximum stress that
the conductor can withstand rather than by the maximum field that it can support. Novel
concepts such as stress-managed coils  will allow this challenge to be addressed.

\subsubsection{Beam induced background}
Muon beam decay produces a significant flux of secondary and tertiary particles in
the detector in spite of the shielding masks.
A detailed study of the impact of this background on the detector performance
has been performed at center of mass of 1.5 TeV using two different Monte
Carlo programs, Fluka and Mars15. The results agree within a factor
less than 2 and demonstrate that the beam-induced background does not limit
the physics potential with the current detector technologies and the
proposed shielding.

A preliminary investigation of the background effects at center of mass of
3 TeV has been performed by using the Fluka simulation and the MAP Interaction
region (IR) designed in 2018 for a 1.5 TeV CM. The shielding structure and the detector are kept exactly the
same as used for the 1.5 TeV. This is not optimal for the 3 TeV Interaction
Region. Several improvements are already being discussed. Under these
conservative assumptions, the results show that the beam-induced background
effects on the detector are similar to those found in the 1.5 TeV study and
the characteristics are determined by the shielding structure. This gives
confidence that the background can also be managed at higher energies.

The collaboration will optimize the configuration of the IR together with
the shielding elements to reduce their dimensions, therefore increasing the
detector acceptance in the forward region. A proper combined optimization of the system of detector, shielding and IR will enable avoidance of background hot spot points
increasing the detector performance in the forward region.
These activities require a very close collaboration among detector and
accelerator experts, who have to work more closely than than in previous facility design activities.

Experts are required in novel shielding materials that can absorb different particle species and momenta. The lessons learned in design of the 3 TeV system will serve as a starting point for the 10 TeV case. Here a change of paradigm is needed. No detector has been designed for such high energy lepton collisions.  The beam-induced background and its shielding, together with a possible detector will be studied by using the tools developed for the 3 TeV case, keeping in mind the change of paradigm.

\subsubsection{Neutrino flux}
The decay of muons in the collider ring produces a dense flux of neutrinos
that will exit from the ground at significant distance from the collider. The flux is very dense because the neutrinos have a very small angle with respect to the initial muon trajectory.

Studies are underway to reduce the density of this flux from the collider ring arcs such that it leads to
a negligible impact on the environment, similar to the LHC.
A solution has been proposed to place beam line
components on movers and deforming the ring periodically in small steps such that the muon beam does not point to a specific location for an extended time. This solution allows to reduce the flux to a negligible level and an order of magnitude below the goal of the MAP study, even for a 14 TeV collider placed 200 m deep underground.
The study will
address the mechanical aspects of the solution and its impact on the beam operation.

A dedicated effort, supported by civil engineers, beam scientists, FLUKA and radiation experts will also assess the impact of the insertions on the neutrino flux.

The impact of accidental beam loss can be mitigated by placing the tunnel sufficiently deep. In this case a lost muon beam would not be able to penetrate the Earth sufficiently to escape from the surface. The required depth is less than for the neutrino flux mitigation.

\subsection {Required R\&D}

The MAP Collaboration initiated its study with an evaluation of the feasibility of the key sub-systems
required to deliver an energy-frontier collider. Several issues were identified as part of the MAP
Feasibility Assessment that had the greatest potential to prevent the realisation of a viable muon collider
concept. These included:

\begin{itemize}
    \item operation of RF cavities in high magnetic fields in the front end and cooling channel;
    \item development of a 6D cooling lattice design consistent with realistic magnet, absorber, and RF cavity specifications;
     \item a direct demonstration and measurement of the ionisation-cooling process;
     \item development of very-high-field solenoids to achieve the emittance goals of the Final Cooling system;
     \item demonstration of fast-ramping magnets to enable RCS capability for acceleration to the TeV scale.

\end{itemize}

While other machine design and engineering conceptual efforts were pursued to develop the overall
definition of a muon collider facility, research in the above feasibility areas received the greatest attention as part of the MAP effort.
An important outcome of MAP was that progress in each of the above areas was sufficient to
suggest that there exists a viable path forward. The test program at Fermilab’s MuCool test area demonstrated operation of gas-filled and vacuum pillbox cavities with up to 50 MV/m gradients in strong
magnetic fields \cite{bib_muon:MAP_GasRF,bib_muon:MAP_VacRF}; a 6D cooling lattice was designed that incorporated reasonable physical assumptions
\cite{bib_muon:MAP_Rectilinear}; a final cooling channel design, which implemented the constraint of a 30 T maximum
solenoid field, came within a factor of two of meeting the transverse emittance goal for a high energy
collider \cite{bib_muon:MAP_FinalCooling};
and while further R\&D is required, fast-ramping magnet concepts \cite{bib_muon:MAP_RampingMags}  do exist that could deliver
TeV muon beams.
Since the end of the MAP studies a number of technologies have developed, which make the
muon collider a promising avenue of study. In particular, new studies are required to leverage the now increased
limits of solenoids and RF cavities, which theory suggests should give an improved cooling
channel performance

\begin{figure}
        \includegraphics[width=0.632\textwidth, trim={2.1cm 0 0 0}]
        {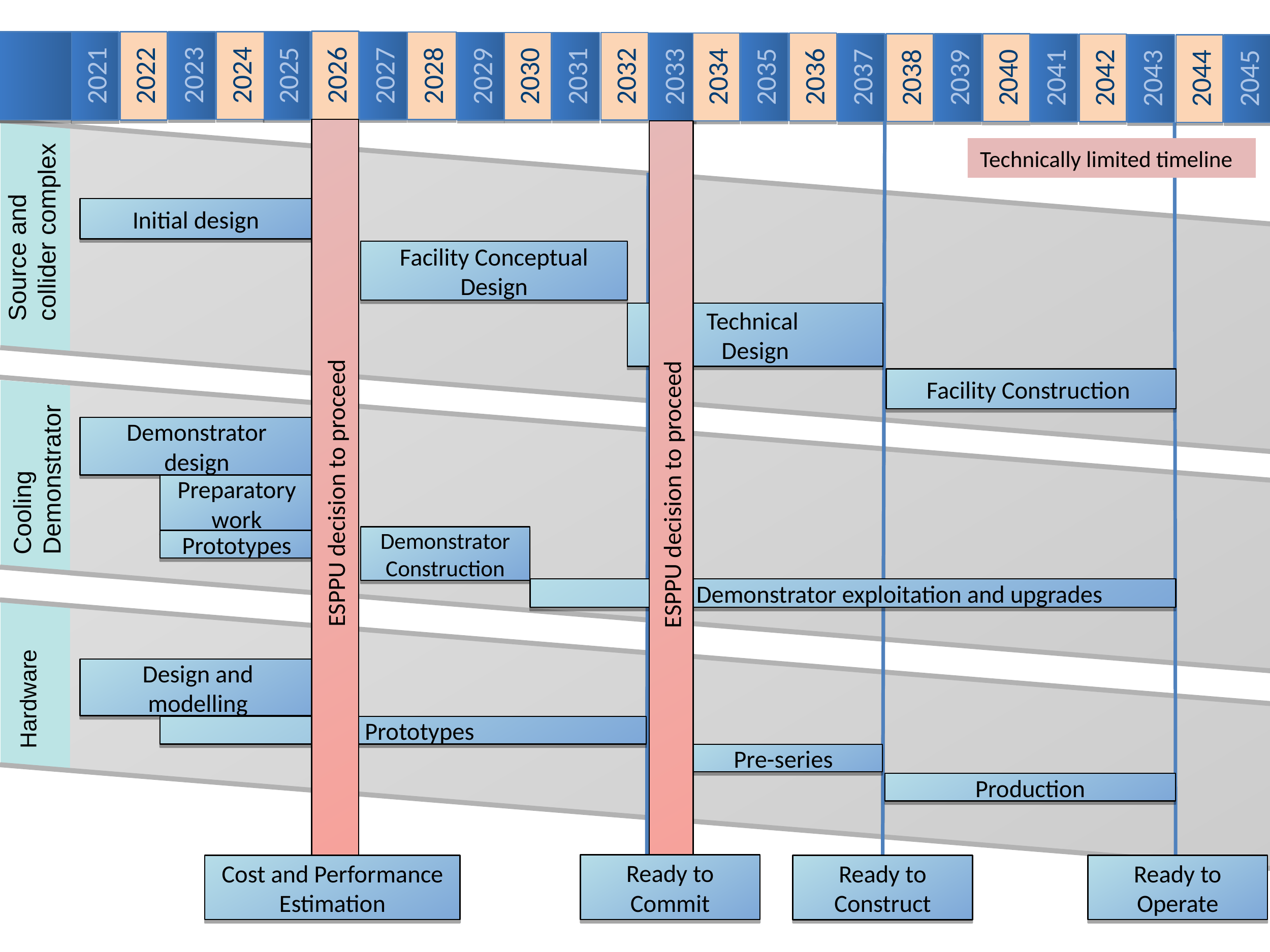}
        \centering\textbf{}

\caption{A technically limited timeline for the Muon Collider R\&D programme}
\label{fig_muon:RDtimeline}
\end{figure}

A muon collider with a centre-of-mass energy around 3 TeV could be delivered on a time scale compatible with the end of operation of the HL-LHC. A technically limited time line is shown in Fig.~\ref{fig_muon:RDtimeline} and discussed in greater detail here \cite{bib_muon:LDGReport}. The muon collider R\&D programme will consist of the initial phase followed by the conceptual and the technical design phases. The initial phase will establish the potential of the muon collider and the required R\&D programme for the subsequent phases. 

The performance and cost of the facility would be established
in detail. A programme of test stands and prototyping of equipment would be performed over a
five-year period, including a cooling cell prototype and the possibility of beam tests in a cooling demonstrator.
This programme is expected to be consistent with the development of high field solenoid and
dipole magnets that could be exploited for both the final stages of cooling and the collider ring development.
A technical design phase would follow in the early 2030s with a continuing programme focusing
on prototyping and preseries development before production for construction begins in the mid-2030s,
to enable delivery of a 3 TeV collider by 2045. The programme is flexible, in order to match the prioritisation
and timescales defined by the next ESPPU, P5 and equivalent processes.

\begin{table*}
\caption{Tentative target parameters for a muon collider at different
  energies based on the MAP design with modifications. These values are only to give a first,
  rough indication. The study will develop coherent parameter sets of its own.}
  
\label{MC:t:param}
\begin{center}
  \begin{tabular}{|*6{c|}}
    \hline
    Parameter & Symbol & Unit &\multicolumn{3}{c|}{Target value} \\
    \hline
    Centre-of-mass energy & $E_\text{cm}$ & \si{\tera\electronvolt} & 3 & 10 & 14 \\
    Luminosity & $\cal L$ & \SI{e34}{\per\square\centi\meter\per\second} & 1.8 & 20 & 40 \\
    Collider circumference& $C_\text{coll}$ & \si{\kilo\meter} & 4.5 & 10 & 14 \\
    \hline
    Muons/bunch & $N$ & \num{e12} & 2.2 & 1.8 & 1.8 \\
    Repetition rate & $f_\text{r}$ & \si{\hertz} & 5 & 5 & 5 \\
    Beam power  & $P_\text{coll}$ & \si{\mega\watt} &5.3  & 14.4 &20 \\
    Longitudinal emittance& $\epsilon_\text{L}$ & \si{\mega\electronvolt\meter} & 7.5 & 7.5 & 7.5 \\
    Transverse emittance& $\epsilon$ & \si{\micro\meter} & 25 & 25 & 25 \\
    \hline
    IP bunch length& $\sigma_z $ & \si{\milli\meter} & 5 & 1.5 & 1.07 \\
    IP beta-function& $\beta $ & \si{\milli\meter} & 5 & 1.5 & 1.07 \\
    IP beam size& $\sigma $ & \si{\micro\meter} & 3 & 0.9 & 0.63 \\
    \hline
  \end{tabular}
\end{center}
\end{table*}

Based on the MAP design, target parameter sets have been defined for the collider as a starting point, shown in table~\ref{MC:t:param} above. The parameter sets have a luminosity-to-beam-power ratio that increases with energy. They are based on using the same muon source for all energies and a limited degradation of transverse and longitudinal emittance with energy. The design of the technical components to achieve this goal are a key element of the study. It is important to emphasize that a 10~TeV lepton collider is uncharted territory and poses a number of key challenges.

\begin{itemize}
    \item The collider can potentially produce a high neutrino flux that might lead to increased levels of radiation far from the collider. This must be mitigated and is a prime concern for the high-energy option.
    \item The Machine Detector Interface (MDI) might limit the physics reach due to beam-induced background, and the detector and machine need to be simultaneously optimised.
    \item The collider ring and the acceleration system that follows the muon cooling can limit the energy reach. These systems have not been studied for 10~TeV or higher energy. The collider ring design impacts the neutrino flux and MDI.
    \item The production of a high-quality muon beam is required to achieve the desired luminosity. Optimisation and improved integration are required to achieve the performance goal, while maintaining low power consumption and cost. The source performance also impacts the high-energy design.
\end{itemize}

Integrated accelerator design of the key systems is essential to evaluate the expected performance, to
validate and refine the performance specifications for the components and to ensure beam stability and
quality.  A description of the key technology challenges and their relation to the state-of-the art can be found here \cite{bib_muon:LDGReport}. 
 
\subsection {Required and Desirable Demonstrators}

Demonstrations are required both for the muon source and the high energy complex. The compact nature of the muon cooling system, high gradients and relatively high-field solenoids present some unique challenges that require demonstration. The high-power target also has a number of challenges that should be evaluated using irradiation facilities or single impact beam tests. The issues in the high energy complex arise from the muon lifetime. Fast acceleration systems and appropriate handling of decay products result in unique challenges for the equipment. The following new facilities are required and will be developed or constructed as part of the program:

\subsubsection{Ionization cooling demonstrator}
MICE only demonstrated transverse cooling without re-acceleration and operated at relatively high emittance. Further tests must be performed to demonstrate the 6D cooling principle at low emittance and including re-acceleration through several cooling cells. Many of the challenges are associated with integration issues of the magnets, absorbers and RF cavities. For example, operation of normal conducting cavities near to superconducting magnets may compromise the cryogenic performance of the magnet. Installation of absorbers, particularly using liquid hydrogen, may be challenging in such compact assemblies. In order to understand and mitigate the associated risks, an offline prototype cooling system will be required. Such a system will require an assembly and testing area, with access to RF power and support services. This could be integrated with the demonstrator facility, as it will need an area for staging and offline testing of equipment prior to installation on the beamline. The possibility to perform intensity studies with a muon beam are limited. In the first instance such effects should be studied using simulation tools. If such studies reveal potential technical issues, beam studies in the presence of a high intensity source will be necessary, for example using protons.

\subsubsection{Ionization Cooling RF development}
The cooling systems require normal-conducting RF cavities that can operate with high gradient in strong magnetic fields without breakdown. No satisfactory theory exists to model the breakdown. Considerable effort was made by MAP to develop high-gradient RF cavities. Two test cavities have been developed: The first cavity was filled with gas at very high pressure. The second cavity used beryllium walls. Both tests presented promising results. In the meantime operation of normal-conducting RF cavities at liquid nitrogen temperature has been demonstrated to reduce multipacting. In order to test the concepts above and others further, a dedicated test facility is required. An RF source having high peak power at the appropriate frequency and a large aperture solenoid that can house the RF cavity will be needed. No such facility exists at present.

\subsubsection{Cooling magnet tests}
In order to improve cooling performance high field magnets are required with opposing-polarity coils very close together. The possibility to implement high-field magnets (including those based on HTS) will be investigated, with appropriate design studies leading to the construction of high-field solenoid magnets having fields in the range 20 T to 25 T. Very high field magnets are required for the final cooling system. In this system, the ultimate transverse emittance is reached using focusing in the highest-field magnets As a first step, a 30 T magnet, corresponding to the MAP baseline, would be designed and constructed. Feasibility studies towards a 50 T magnet would also be desirable, which may include material electro-mechanical characterization at very high field as well as technology demonstration at reduced scale. These very demanding magnets are envisaged to be developed separately to the cooling demonstrator. Eventually they could be tested in beam if it was felt to be a valuable addition to the programme. In order to support this magnet R\&D, appropriate facilities will be required. Testing of conductors requires a suitable test installation, comprising high field magnets, variable temperature cryogenics and high-current power supplies. Magnet development and test will also require these facilities in addition to access to appropriate coil and magnet manufacturing capabilities.

\subsubsection{Acceleration RCS magnets}
Acceleration within the muon lifetime is rather demanding. The baseline calls for magnets that can cycle through several T on a time scale of a few ms. The exact specification will be defined during the design work, but it is clear that a resonant circuit is the best suited solution to power the magnets. The design of the magnet and powering system will be highly integrated, and work on scaled prototypes is anticipated. Superconducting RCS magnets may offer higher field reach than normal-conducting magnets, but are challenging to realise owing to heating arising from energy dissipation in the conductor during cycling (AC loss). This heating can lead to demands on the cryogenic systems that outweigh the benefits over normal-conducting magnets. Recent prototypes have been developed using HTS that can operate at higher temperatures, and in configurations leading to lower AC losses, yielding improved performance. In order to continue this research, magnet tests with rapid pulsed power supplies and cryogenic infrastructure will be required.
\subsubsection{Effects of radiation in material}
The high beam power incident on the target and its surroundings is very demanding. Practical experience from existing facilities coupled with numerical studies indicate that there will be challenges in terms of target temperature and lifetime. Instantaneous shock load on the target will also be challenging. Tests are foreseen to study behaviour of target material under beam in this instance. Tests are desirable both for instantaneous shock load and target lifetime studies.
Additionally, the effect of radiation in the target
region on the superconducting materials (LTS and HTS) and insulators is an important parameter. As the target solenoid design matures, additional studies may be required taking into account the magnet arrangement, conductor design and estimates of radiation levels. In order to realise such tests, facilities having both instantaneous power and integrated protons on target equivalent to the proton beam parameters assumed for this study are desirable. The database of radiation effects on superconductors (HTS) and insulators also requires an extension to cover the projected conditions in the target area.
\subsubsection{Superconducting rf cavities}
Development of efficient superconducting RF with large accelerating gradient is essential for the high energy complex. Initially work will focus on cavity design; however eventually a high gradient prototype at 300-400 MHz frequency will be required. In order to realise such a device, appropriate superconducting cavity production and test facilities will be required including surface preparation techniques and a capability for high power tests.

\section {Staging Options and Upgrades}
At this moment no limit is known to the energy reach of the muon collider.
Practical limits such as cost and power consumption as well as the luminosity
reach at each energy are the key considerations for the choice of energies.
Other limits might come from the impact on the site resulting from the
neutrino flux or the background conditions in the experiments since they
could also limit the luminosity.

We therefore consider a programme with energy stages to limit the cost and risk
of each stage. Currently, there is no obvious strong benefit of reducing the
luminosity at each energy stage and foresee future large upgrades. However,
smaller scale improvements of the luminosity at each energy stage can be
implemented as will be discussed below.

\subsection{Energy Upgrades}
The muon collider can be implemented as a staged concept providing a road
toward higher energies. The accelerator chain can be expanded by an additional
accelerator ring for each energy stage. A new collider ring is required for
each energy stage. Currently, we envisage that the new accelerator ring would
be a hybrid pulsed synchrotron that uses a combination of fast-ramping magnets
interleaved with static superconducting magnets. Hence it may be possible to
to reuse the magnets and other equipment of the previous collider ring.

Currently, the focus of the study is on 10 TeV. This energy is well beyond the
3 TeV of the third and highest energy stage of CLIC, the highest energy
proposal with a CDR. An potential intermediate energy of 3 TeV is envisaged at
this moment. Its physics case is similar to the final stage of CLIC and it is
expected that this stage would roughly cost half as much as the 10 TeV stage.

A 3 TeV stage is less demanding in several technological areas. It may not be
necessary to implement a mechanical neutrino flux mitigation system in the
collider ring arcs, moving the beam inside of the magnet apertures may be
sufficient. The final focus magnets require an aperture and a gradient
comparable to the values for HL-LHC.

In general, it will be possible to implement larger margins in the design
at 3 TeV. The operational experience will then allow to accept smaller margins
at 10 TeV.

The strength of its dipoles drives the collider ring size and impacts the
luminosity. The cost optimum is given by the magnet and tunnel cost; it is
possible that cheaper, well stablished magnet technology ~-~ such as
NbTi at this moment ~-~ would result in a lower cost even if the tunnel has to
be longer. For fixed beam current, the luminosity is proportional to the
magnet field and is one aspect of the trade-off.

Once the cost scale of the muon collider concept is more precisely known
and once the mitigation of the technical challenges are well defined,
the energy staging will be reviewed taking into account the physics case
and additional considerations from the site and reuse of existing equipment
and infrastructure, such as the LHC tunnel. This will provide a sustainable
road with important physics cases for its affordable stages.

\subsection{Luminosity Upgrades}
After the construction of each energy stage the luminosity can be increased
by limited modifications using improved technologies. This includes
in particular the final cooling and the final focusing systems.

The luminosity of the muon collider is driven by the beam power and brightness
and the ability to take advantage of this brightness by focusing
to very small sizes beam sizes in the collision point.
The beam power is closely linked to the power consumption of the collider and
depends on the efficiency of the muon cooling complex, the accelerating RF
systems, the fast-ramping magnet systems and the mitigation of heat load in
the superconducting components of the complex induced by muon decay.

An increase in beam power by increasing the repetition rate can be considered.
In addition to increased power consumption for the RF, fast-ramping magnet
and cooling systems, this would also increase the power in the proton target
which is expected to be an important limitation. Whether an operation of
two targets by alternating the pulses between them is feasible at acceptable
cost remains to be evaluated.

A larger beam brightness resulting from an increase of the bunch charge within
the same emittance would also allow to increase the luminosity, even if the
repetition rate would have to be reduced to remain at the same power
consumption. However, this case might have to mitigate intensity bottlenecks
along the collider chain.

The beam brightness depends in particular
on the final muon cooling system. Further improvement of the design and the
technology of this complex are mainly a technical challenge and could be
implemented at relatively limited cost when the muon collider is already
operating. Similarly, one can envisage that better final focus quadrupoles
would allow an increase of the luminosity at a later stage with limited
cost.

\subsection{Experimental system Upgrades}

Several instrumentation challenges need to be addressed in building detectors able to successfully deliver the physics potential of a multi-TeV muon collider.
Detectors and event reconstruction techniques need to be designed to cope with the presence of the beam-induced background (BIB) produced by the muon decays and their interaction with the machine elements at the Machine Detector Interface.  The available and incoming detector technologies were proved suited to be exploited in an experiment at a muon collider operating at $\sqrt{s}=3$~TeV, while the energy regime of the decay products in the environment at $\sqrt{s}=10$~TeV still demands for in depth dedicated studies. Operating conditions expected at $\sqrt{s} = 1.5$~TeV, simulated with highly comparable results both in MARS15 and FLUKA frameworks, for a 200 days of operation per year, result in  $\sim 10^{14-15}$~cm$^{-2}$y$^{-1}$ 1-MeV-neq fluence in the region of the tracking detector and $\sim 10^{14}$~cm$^{-2}$y$^{-1}$ in the electromagnetic calorimeter, with a steeply decreasing radial dependence beyond it. The total ionizing dose is $\sim10^{-3}$~Grad/y on the tracking system and $\sim10^{-4}$~Grad/y on the electromagnetic calorimeter. With the FLUKA simulation at $\sqrt{s}=3$~TeV and the preliminary results obtained with the first lattice at $\sqrt{s}=10$~TeV there is no indication of an effective degradation of the radiation levels in the detector volume. Inner tracking has to cope with a hit density of up to 1,000 hits/cm$^2$, with a  the density decreasing rapidly as a function of the radial distance from the beam-line. In the expected operating conditions, high performance tracking is necessary to achieve good efficiency and resolution for reconstructing charged leptons, jets, energy sums, displaced vertices originating from the heavy flavor jets, as well as potential new phenomena. Promising tracking technologies rely on the high granularity of silicon pixels, necessary to reduce hit occupancy level of a few \%. Moreover excellent timing resolution in  the tracking detector and the calorimeter systems would allow a more efficient background rejection on detector and at filtering level. New dedicated muon spectrometers, trigger systems and data acquisition are under study.
 
It should be mentioned that development of these technologies is beneficial for other future collider options, in particular high energy hadron machines.

\section {Synergies with other concepts and/or existing facilities}
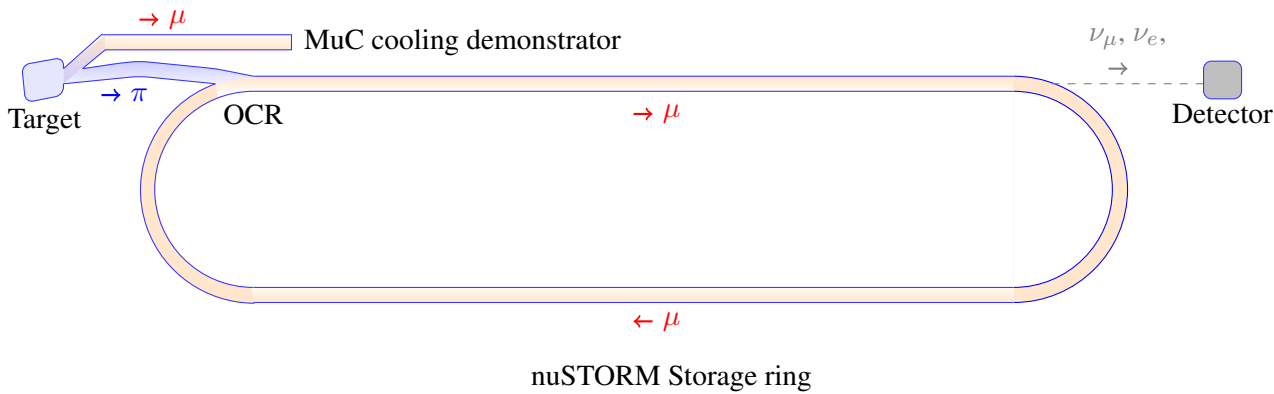
\begin{figure}
\begin{tikzpicture}[scale=0.05]
% target
\draw[blue!80, rounded corners=3pt, fill=blue!10] (10,-2) -- (10,-5) -- (0,-7) -- (-1,3) -- (10,5) -- (10, 2);
\node[] at (5,-12) {Target};

%Demo
\shade[bottom color=blue!20,top color=orange!20] (10,-2) -- (21,7) -- (21,11) -- (10,2);
\shade[bottom color=orange!20,top color=orange!5] (70,11) -- (70,7) -- (21, 7) -- (21, 11);
\draw[blue!80] (10,2) -- (20,11) -- (70,11) -- (70,7) -- (21, 7) -- (15, 2);
\node[] at (115,9) {MuC cooling demonstrator};
\draw[thick,->, color=red] (30,15) -- (35,15);
\node[text=red] at (40,15) {$\mu$};

% storage ring
\shade[top color=orange!20, bottom color=orange!5] (50, 0) -- (260,0) -- (260,-4) -- (50, -4);
\draw[blue!80, fill=orange!20] (50,-2) arc (110:270:30);
\draw[blue!80, fill=white] (60,-4) arc (90:270:26);
\draw[blue!80] (60, 0) -- (260,0);
\draw[blue!80] (60, -4) -- (260,-4);
\draw[blue, fill=orange!20] (260,-60) arc (-90:90:30);
\draw[blue, fill=white] (260,-56) arc (-90:90:26);
\shade[top color=orange!5, bottom color=orange!20] (60, -56) -- (260,-56) -- (260,-60) -- (60, -60);
\draw[blue!80] (60, -56) -- (260,-56);
\draw[blue!80] (60, -60) -- (260,-60);
\draw[thick,->, color=red] (160,-10) -- (165,-10);
\node[text=red] at (170,-10) {$\mu$};
\draw[thick,<-, color=red] (160,-65) -- (165,-65);
\node[text=red] at (170,-65) {$\mu$};
\node[] at (170,-80) {nuSTORM Storage ring};

%Pion line
\shade[bottom color=blue!5,top color=blue!20] (15, 2) -- (25,3.5) -- (30,4) -- (35,3.5) -- (50,2) -- (60,0) -- (50,-2) -- (35,-0.5) -- (30,0) -- (25,-0.5) -- (10, -2);
\draw[blue!80, rounded corners=3pt] (15, 2) -- (25,3.5) -- (30,4) -- (35,3.5) -- (50,2) -- (60,0);
\draw[blue!80, rounded corners=3pt] (10, -2) -- (25,-0.5) -- (30,0) -- (35,-0.5) -- (50,-2);
\draw[thick,->, color=blue] (20,-5) -- (25,-5);
\node[text=blue] at (30,-5) {$\pi$};
\node[] at (60,-10) {OCR};

% detector
\draw[gray, dashed] (270,-2) -- (310, -2);
\draw[blue!80, fill=gray!50, rounded corners=3pt] (310,-6) rectangle (320,4);
\draw[thick,->, color=gray] (285,2) -- (290,2);
\node[text=gray] at (290,10) {$\nu_{\mu}$, $\nu_{e}$, };
\node[] at (315,-10) {Detector};

\end{tikzpicture}
\caption{Schematic showing nuSTORM including the muon cooling demonstrator for the muon collider. \label{fig_muon:nustorm}}
\end{figure}
%\subsection{Synergies on machine technologies} 
The ambitious programme of R\&D necessary to deliver the muon collider has the potential to enhance the science that can be done at other muon-beam facilities. On the contrary, the progresses in other accelerator facilities will also benefit the design and construction of the muon collider in the future.
nuSTORM and ENUBET offer world-leading precision in the measurement of neutrino cross sections and exquisite sensitivity to sterile neutrinos and physics beyond the Standard Model. nuSTORM in particular will require capture and storage of a high-power pion and muon beam and management of the resultant radiation near to superconducting magnets. The target and capture system for nuSTORM and ENUBET may also provide a testing ground for the technologies required at the muon collider and as a possible source of beams for the essential 6D cooling-demonstration experiment, for example as in the schematic shown in fig. \ref{fig_muon:nustorm}. 
The ongoing LBNF and T2HK projects and their future upgrades will develop graphite targets to sustain the bombardment of MW-level proton beams, which may also lead to a solution for the muon production target for the muon collider.  
The next generation searches for charged lepton flavour violation exploit high-power proton beams impinging on a solid target placed within a high-field solenoid, such as Mu2e at FNAL and COMET at J-PARC. The technological issues of target and muon capture for these experiments are similar to those present in the muon collider design.
 The potential to deliver high quality muon beams could enhance the capabilities of muon sources such as those at PSI, J-PARC and ISIS. The use of frictional cooling to deliver ultra-cold positive and negative   muon beams is under study at PSI and may be applicable to the muon collider.
FFAs have been proposed as a route to attain high proton beam power for secondary particle sources such as neutron spallation sources, owing to the potential for high repetition rate and lower wall plug power compared to other accelerator schemes. The technology related to FFA will be beneficial to choose the fast acceleration scheme of the lower-energy acceleration stages of the muon collider. 
High-power proton accelerators are in fast development use throughout the world, including linacs, rapid cycling synchrotrons, and accumulator or compressor rings, for example, proton drivers ranging from hundreds of kW to multiple MW for spallation neutron sources such as SNS, J-PARC, ESS, ISIS and CSNS, those in about 1 MW for neutrino beams such as FNAL and J-PARC proton accelerator complexes, and those in multiple MW and CW mode for accelerator-driven systems such as CiADS and MYRRHA. The progresses made in the facilities will accumulate the required accelerator technologies for the muon collider. 
The underlying technologies required for the muon collider are also of interest in many scientific fields. The delivery of high field solenoid magnets is of great interest to fields as wide ranging as particle physics, accelerator science and imaging technology. Operation of RF cavities with high gradient is of interest to the accelerator community.

%\subsection{Synergies on conventional facilities and green power} 

%\subsection{Synergies for physics research} 

%\cite{bib_muon:ldg2020}

\section{Conclusion}
The muon collider presents enormous potential for fundamental physics research at the energy frontier.
Previous studies,  in particular the MAP study,  have demonstrated feasibility of many critical components  of the facility.  A number of proof-of-principle experiments and component tests, such as MICE,
EMMA and the MuCool RF programme, have been carried out to practically demonstrate the underlying
technologies.

The muon collider is based on novel concepts and is not as mature as the other high-energy lepton
collider options, in particular also compared to the highest energy option CLIC. However, it promises a
unique opportunity to deliver physics reach at the energy frontier on a cost, power consumption and time
scale that might improve significantly on other energy-frontier colliders. At this stage, building
on significant prior work, it was not identified any showstopper in the concept.
Therefore a development path can address the major challenges and deliver
a 3 TeV muon collider by 2045. 

A global effort has identified the R\&D effort that it considers essential to
address these challenges before the next regional strategy processes to a level that allows estimation of the performance,
cost and power consumption with adequate certainty.   Execution of this R\&D is required in order to
maintain the timescale described in this document.  Ongoing developments in underlying technologies
will be exploited as they arise in order to ensure the best possible performance.  This R\&D effort will
allow future strategy processes to make fully informed recommendations.  Based on the subsequent decisions, a significant ramp-up of resources could
be made to accomplish construction by 2045 and exploit the enormous potential of the muon collider.

Bright muon beams are also the basis of neutrino physics facilities such as NuSTORM and ENU-
BET. These could potentially share an important part of the complex with a muon cooling demonstrator.

\bibliographystyle{plain}

\bibliography{main}
%\bibliography{Muon}

\end{document}